\documentclass[aps,nofootinbib,floatfix,showpacs]{revtex4}
\usepackage{mathrsfs}
\usepackage{graphicx}
\usepackage{amsmath}
\usepackage{amsfonts}
\usepackage{amssymb}
\usepackage{color}
\usepackage{lineno}
\usepackage{subfig}
\usepackage{booktabs}
\usepackage{makecell}
\usepackage{multirow}
\usepackage{color}
\usepackage{epsfig}
\usepackage{CJK}
\usepackage{epsfig}
\usepackage{eepic}
\usepackage{bbm}
\usepackage{dcolumn}% Align table columns on decimal point
\usepackage{bm}% bold math
\usepackage{ulem}
\usepackage{multirow}% for the tabular, added in 30.01.2013
\usepackage{slashed}

\newcommand{\omits}[1]{}
\def\bc{\begin{center}}

\def\ec{\end{center}}
\def\be{\begin{eqnarray}}
\def\ee{\end{eqnarray}}

\definecolor{dyellow}{rgb}{1.,0.8,.0}
\definecolor{myblue}{rgb}{.1,.1,.7}
\definecolor{dcyan}{rgb}{.0,.6,.6}
\definecolor{cyan}{rgb}{0.4,1.0,1.0}
\definecolor{dmagenta}{rgb}{0.6,0.0,0.6}
\definecolor{brown}{rgb}{0.6,0.2,0.}
\definecolor{darkblue}{rgb}{.0,.0,0.5}
\definecolor{darkred}{rgb}{0.75,0.0,0.0}
\definecolor{orange}{rgb}{1.,.6,.0}
\definecolor{dorange}{rgb}{0.8,.4,.0}
\definecolor{green}{rgb}{0.0,1.0,0.0}
\definecolor{darkgreen}{rgb}{0.0,0.6,0.0}
\definecolor{purple}{rgb}{.4,.0,.4}
\definecolor{lightgrey}{rgb}{0.7, 0.7, 0.7}
\definecolor{grey}{rgb}{0.4, 0.4, 0.4}
\begin{document}

%\preprint{hep-th/yymmnnn}

\title{Topology of black hole thermodynamics via R{\'e}nyi statistics}

\author{Chen-Wei Tong$^1$} \email{tongchw3@mail2.sysu.edu.cn}
\author{Bin-Hao Wang$^{2}$} \email{202121213@stumail.nwu.edu.cn}
\author{Jia-Rui Sun$^{1}$} \email{sunjiarui@mail.sysu.edu.cn}

\affiliation{${}^1$School of Physics and Astronomy, Sun Yat-Sen University, Guangzhou 510275, China}
\affiliation{${}^2$School of Physics, Northwest University, Xi’an 710127, China}
%\date{\today}

%% REVTEX4
%\maketitle

\begin{abstract}
In this paper, we investigate the topological numbers of the four-dimensional Schwarzschild black hole, $d$-dimensional Reissner-Nordstr\"{o}m (RN) black hole, $d$-dimensional singly rotating Kerr black hole and five-dimensional Gauss-Bonnet black hole via the R{\'e}nyi statistics. We find that the topological number calculated via the R{\'e}nyi statistics is different from that obtained from the Gibbs-Boltzmann (GB) statistics. However, what is interesting is that the topological classifications of different black holes are consistent in both the R{\'e}nyi and GB statistics: the four-dimensional RN black hole, four-dimensional and five-dimensional singly rotating Kerr black holes, five-dimensional charged and uncharged Gauss-Bonnet black holes belong to the same kind of topological class, and the four-dimensional Schwarzschild black hole and $d(>5)$-dimensional singly rotating Kerr black holes belong to another kind of topological class. In addition, our results suggest that the topological numbers calculated via the R{\'e}nyi statistics in asymptotically flat spacetime background are equal to those calculated from the standard GB statistics in asymptotically AdS spacetime background, which provides more evidence for the connection between the nonextensivity of the R{\'e}nyi parameter $\lambda$ and the cosmological constant $\Lambda$.
\end{abstract}

%% REVTEX4
%\pacs{04.62.+v, 04.70.Dy, 12.20.-m}

\maketitle
%\tableofcontents

%%%%%%%%%%%%%%%%%%%%%%%%%%%%%%%%%%%%%%%%%%%%%%%%%%%%%%%%%%%%%%%%%%%%%%
\section{Introduction}
%%%%%%%%%%%%%%%%%%%%%%%%%%%%%%%%%%%%%%%%%%%%%%%%%%%%%%%%%%%%%%%%%%%%%%
\label{introduction}
The studying of thermodynamics in gravitational systems has made great progresses in revealing the important properties of black holes~\cite{Bardeen:1973gs,Bekenstein:1973ur,Hawking:1976de}. For example, one of the most important thermodynamic aspects in Anti-de Sitter (AdS) spacetime is the Hawking-Page (HP) phase transition between thermal radiation and large AdS black hole~\cite{Hawking:1982dh}, which can be interpreted as the confinement/deconfinement phase transition of the dual boundary quantum fields in the context of AdS/CFT correspondence~\cite{Maldacena:1997re,Witten:1998qj,Witten:1998zw,Birmingham:2002ph}. And for the charged AdS black hole, there exists a first order phase transition between the charged small and large black holes, which is similar to the van der Waals phase transition~\cite{Chamblin:1999tk,Chamblin:1999hg}. In addition, new thermodynamic perspective with the extend phase space can be introduced for black holes in the presence of the cosmological constant $\Lambda$, in which $\Lambda$ is regarded as the thermodynamic pressure $P$~\cite{Kastor:2009wy,Dolan:2010ha,Dolan:2011xt}, see also \cite{Kubiznak:2012wp,Wei:2012ui,Cai:2013qga,Wei:2014hba,Cai:2014jea,Cai:2014znn,Wei:2015iwa,Zhang:2015ova,Caceres:2015vsa,Kubiznak:2016qmn,Ghosh:2019pwy,Xu:2020gud,Xu:2021qyw} for related studies.

Recently, Wei, Liu and Mann~\cite{Wei:2022dzw} proposed a new approach for describing black hole thermodynamics by using Duan's topological current $\phi$-mapping theory~\cite{duan1984structure,Duan:1979ucg}. It is shown in this approach that a black hole can be regarded as defects in thermodynamic parameter space, from the topological perspective, the winding numbers can reflect the characteristics of local thermodynamic stability of the black hole, and the topological number defined as the sum of winding numbers can divide different black hole solutions into three categories. Subsequently, the method has been applied to reanalyze thermodynamic properties of various black holes by calculating their topological numbers, see for example~\cite{Yerra:2022coh,Bai:2022klw,Liu:2022aqt,Fan:2022bsq,Fang:2022rsb,Wu:2022whe,Wu:2023sue,Wu:2023xpq,
	Du:2023wwg,Fairoos:2023jvw,Gogoi:2023xzy,Yerra:2023hui,Zhang:2023tlq,Hung:2023ggz,Wu:2023fcw,Sadeghi:2023aii,
	Chen:2023ddv}.

On the other hand, as for the thermodynamic description of black holes, there still exist some outstanding issues that deserve further studying. It is known that the Bekenstein-Hawking entropy of black hole is nonextensive which is proportional to the area of the horizon rather than the volume. Therefore, the standard Gibbs-Boltzmann (GB) statistics may not be the best choice in characterizing strongly gravitating systems. In other words, the GB entropy formula, which satisfies the condition of neglecting long-range type interactions, is violated for the long-range interactions (e.g. black hole system)~\cite{gibbs1902elementary,Tsallis:2012js}. Moreover, based on the GB statistics, it is generally believed that a black hole in asymptotically flat spacetime has a negative heat capacity, which implies that the thermodynamic system of black hole can not be in thermodynamic equilibrium with a heat bath of thermal radiation, and the canonical ensemble in the GB statistics may not reliable in nonextensive long-range interaction black hole system. Hence it is necessary to find more appropriate statistical approach to describe systems with long-range type interactions. In general, for a non-additive system, the Abe's composition rule \cite{abe2001general} is given by
\begin{equation}\label{Abe}
	H_{\lambda }(S_{12})=H_{\lambda }(S_{1})+H_{\lambda }(S_{2})+\lambda H_{\lambda }(S_{1})H_{\lambda }(S_{2}),
\end{equation}
where $H_{\lambda}$ is a differentiable function of $S$ and $\lambda \in \mathbb{R} $ is a constant parameter. A simple example is the Tsallis entropy~\cite{Tsallis:1987eu}, which has the following form
\begin{equation}\label{Tsallis}
	S_{\rm T}=\frac{1}{1-q}(\sum_{i=1}^{N} p_{i}^{q}-1), \quad q\in \mathbb{R}
\end{equation}
where $N$ is the total number of microstates and $p_{i}$ are the probabilities of system. When $q \to 1$, $S_{\rm T} \to S_{\rm GB}=-\sum_{i=1}^N p_{i} \ln p_{i}$, which is the standard GB statistics. One can find that the composition rule of Tsallis entropy satisfies Eq.(\ref{Abe}), i.e.,
\begin{equation}
	S_{\rm T}(A,B)=S_{\rm T}(A)+S_{\rm T}(B)+(1-q)S_{\rm T}(A)S_{\rm T}(B),
\end{equation}
However, there is \textcolor{red}{an} incompatibility between the zeroth law of thermodynamics and non-additive entropy composition rules in nonextensive thermodynamics. Fortunately, this problem can be solved by the so-called the formal logarithm approach proposed by Bir{\'o} and V{\'a}n ~\cite{PhysRevE.83.061147}. For the homogeneous system, the Tsallis entropy can be transformed into zeroth law compatible entropy function as
\begin{equation}\label{transrule}
	L\left(S_{\rm T}\right)=\frac{1}{1-q}\left[\ln \left(1+(1-q) S_{\rm T}\right)\right] \equiv S_{\rm R} ,
\end{equation}
which is just the well-known R{\'e}nyi entropy \cite{renyi1959dimension} and satisfies the additive relation
\begin{equation}
	S_{\rm R}(A,B)=S_{\rm R}(A)+S_{\rm R}(B).
\end{equation}
Therefore, the corresponding zeroth law compatible temperature function can be calculated as
\begin{equation}
	\frac{1}{T_{\rm R}}=\frac{\partial S_{\rm R}(E)}{\partial E},
\end{equation}
where $E$ is the energy of the system. It is interesting to regard $S_{\rm{BH}}$ as $S_{\rm T}$ \cite{Tsallis:2012js}, and it is natural to apply the above approach to investigate the black hole entropy \cite{Biro:2013cra}. In recent years, the R{\'e}nyi statistics had been applied into different kinds of black holes \cite{Czinner:2017tjq,Promsiri:2020jga,Abreu:2020vkc,Promsiri:2021hhv,Barzi:2022ygr,
	ElMoumni:2022chi,Hirunsirisawat:2022fsb,Wang:2023lmr,Barzi:2023mit}, in which the parameter $\lambda$ is shown to play a role of pressure just like the cosmological constant $\Lambda$.

As discussed above, the analysis of topological number for various kinds of black holes is mainly based on GB statistics. It is interesting to extend the thermodynamic topology to the black holes via the R{\'e}nyi statistics. In the present paper, we will combine topological method with the R{\'e}nyi statistics to study the following two problems: The first one is that the topological number is expected to change due to the presence of nonextensivity parameter $\lambda$, but whether the calculation of topological number via R{\'e}nyi statistics will change the topological classification of the black hole. The second one is that, previous studies have shown evidences supporting the proposal that there exists \textcolor{red}{a} relation between nonextensivity parameter $\lambda$ and the cosmological constant $\Lambda$, it is worth considering whether the topological numbers in asymptotically flat and asymptotically AdS spacetime calculated via R{\'e}nyi and GB statistics are also related, respectively.

The outline of the paper is as follows. In Sec. \ref{topology}, we give a brief review of the topological approach. In Sec. \ref{Schwarzschild topology}, we start with four-dimensional Schwarzschild black hole and calculate its topological number via the R{\'e}nyi statistics. In Sec. \ref{RN topology} and \ref{Kerr topology}, we calculate the topological numbers of Reissner-Nordstr{\"o}m and Kerr singly rotating black holes in four and higher dimensions, respectively. In Sec. \ref{5d GB black hole}, we analyze the topological number of five-dimensional charged black hole in Gauss–Bonnet gravity. Finally, we give the discussion and conclusion in Sec. \ref{discussion}.

%%%%%%%%%%%%%%%%%%%%%%%%%%%%%%%%%%%%%%%%%%%%%%%%%%%%%%%%%%%%%%%%%%%%%%
\section{Topology of black hole thermodynamics}	\label{topology}
%%%%%%%%%%%%%%%%%%%%%%%%%%%%%%%%%%%%%%%%%%%%%%%%%%%%%%%%%%%%%%%%%%%%%%

In this section, we first give a brief review of the topological approach. The generalized off-shell free energy of a black hole can be written as \cite{Wei:2022dzw}
\begin{equation}
	\mathcal{F}=M-\frac{S}{\tau},
\end{equation}
where $M$ and $S$ are the black hole mass and entropy, respectively, $\tau$ is a variable which can be regarded as the inverse temperature of the ensemble. Only when $\tau = \tau_{\rm H}=1/T_{\rm H}$, where $T_{\rm H}$ is the Hawking temperature, the generalized free energy is on-shell and reduces to the Helmholtz free energy $\mathcal{F}= M-T_{\rm H}S$. In terms of the R{\'e}nyi statistics, the generalized off-shell free energy can be rewritten as follows
\begin{equation}
	\label{FR}
	\mathcal{F}_{\rm R}=M-\frac{S_{\rm R}}{\tau_{\rm R}}.
\end{equation}
The generalized free energy becomes on-shell and reduces to $\mathcal{F}_{\rm R}=M - T_{\rm R}S_{\rm R}$ when $\tau_{\rm R} = 1/T_{\rm R}$ (In the following section, if there is no special explanation, the generalized free energy we refer to indicates the one in Eq.(\ref{FR}).). Then a vector $\phi$ can be introduced as
\begin{equation}
	\phi=\left(\frac{\partial \mathcal{F}_{\rm R}}{\partial r_{\rm h}},-\cot \Theta \csc \Theta\right),
\end{equation}
where $0< r_{\rm h}<+ \infty$ and $0\le \Theta \le \pi$. When $\Theta= 0 ,\pi$, $\phi^{\Theta}$ is divergent and the direction of the vector points outward. Further, by using Duan's $\phi$-mapping topological current theory \cite{Duan:1979ucg,Duan:1998kw,Fu:2000pb}, a topological current can be defined as
\begin{equation}
	j^{\mu}=\frac{1}{2 \pi} \epsilon^{\mu \nu \rho} \epsilon_{a b} \partial_{\nu} n^{a} \partial_{\rho} n^{b}, \quad \mu, \nu, \rho=0,1,2
\end{equation}
where $\partial_{\nu}=\frac{\partial}{\partial x^{\nu}}$ and $x^{\nu}=(\tau_{\rm R},\ r_{\rm h},\ \Theta)$. The unit vector $n$ is given by $n=(n^{1}, n^{2})$, where $n^{1}=\phi^{1}/\parallel \phi\parallel $ and $n^{2}=\phi^{2}/\parallel \phi\parallel $. It is easy to check that the topological current is conserved, i.e., $\partial_{\mu}j^{\mu}=0$. Following the $\phi$-mapping theory, the topological current $j^{\mu}$ can be re-expressed as
\begin{equation}
	j^{\mu}=\delta^{2}(\phi) J^{\mu}(\frac{\phi}{x}),
\end{equation}
where the vector Jacobi is defined as
\begin{equation}
	\epsilon^{a b}J^{\mu}(\frac{\phi}{x})=\epsilon^{\mu\nu\rho}\partial_{\nu}\phi^{a}\partial_{\rho}\phi^{b}.
\end{equation}
Obviously, the topological current $j^{\mu}$ is nonzero only when $\phi=0$, and we can find that the topological charge (or topological number ) can be derived as
\begin{equation}
	W=\int_{\Sigma} j^{0} d^{2} x=\sum_{i=1}^{N} \beta_{i} \eta_{i}=\sum_{i=1}^{N} w_{i},
\end{equation}
where $\beta_{i}$ is the positive Hopf index, which counts the number of the loops of the vector $\phi^{a}$ in the $\phi$ space when $x^{\mu}$ goes around the zero point $z_{i}$, the Brouwer degree $\eta_{i}=$ sign$(J^{0}(\phi/x)_{z_{i}})=\pm 1$ and $w_{i}$ is the winding number for the $i$-th zero point of $\phi$ in the parameter space $\Sigma$. Note that when we choose $\Sigma$ to be the neighborhood of a zero point of $\phi$, it will display local topological properties, on the contrary, if $\Sigma$ is the entire parameter space, the global topological $W$ number will be revealed.

%%%%%%%%%%%%%%%%%%%%%%%%%%%%%%%%%%%%%%%%%%%%%%%%%%%%%%%%%%%%%%%%%%%%%%
\section{The topological number of four-dimensional Schwarzschild black hole}\label{Schwarzschild topology}
%%%%%%%%%%%%%%%%%%%%%%%%%%%%%%%%%%%%%%%%%%%%%%%%%%%%%%%%%%%%%%%%%%%%%%

\begin{figure}[htbp]
	\centering
	\subfloat[]{\includegraphics[width=0.48\columnwidth,height=0.48\linewidth]{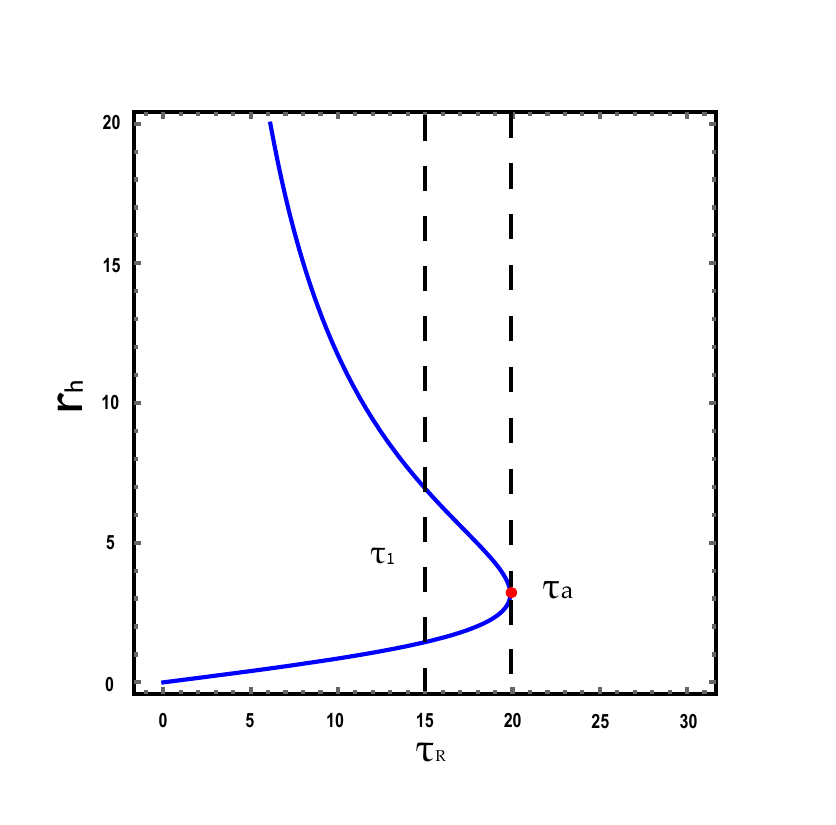}\label{fig.1-1}}
	\quad
	\subfloat[]{\includegraphics[width=0.48\columnwidth,height=0.48\linewidth]{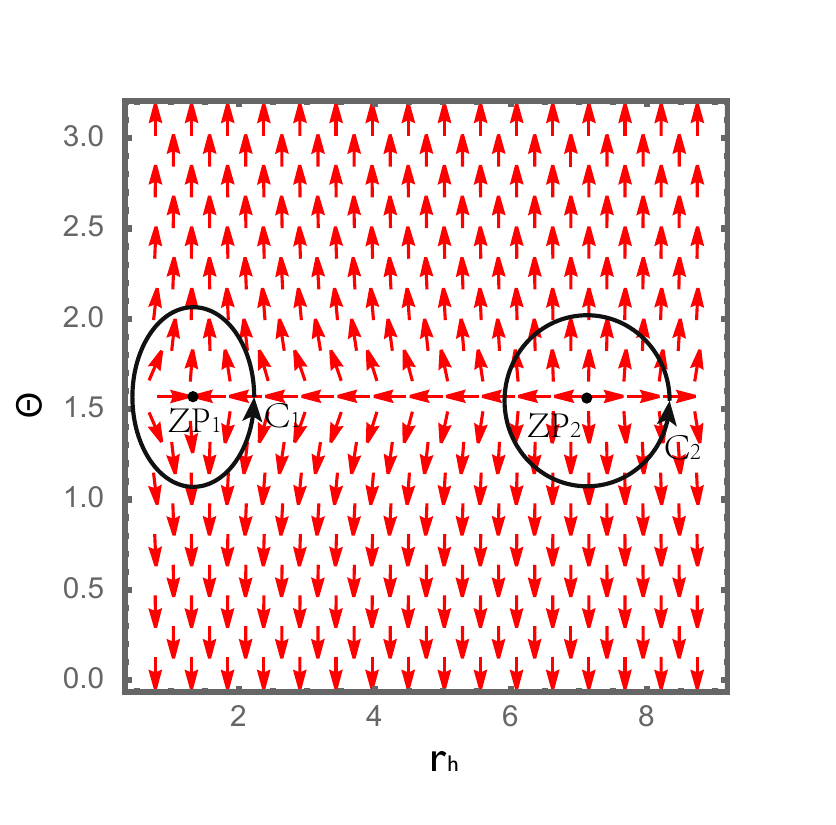}\label{fig.1-2}}
	
	\caption{The left figure (a): zero points of the vector $\phi^{r_{\rm h}}$ shown in the $r_{\rm h}-\tau_{\rm R}$ plane with $\lambda= 0.1/\pi $ for four-dimensional Schwarzschild black hole. The annihilation point for this black hole is represented by the red dot with $\tau_{a}$.
		There are two Schwarzschild black holes when $\tau_{\rm R}=\tau_{1}$. Obviously, the topological number is: $W =-1+1 = 0$. The left figure (b): the red arrows represent the unit vector field $n$ on a portion of the $r_{\rm h}-\Theta$ plane with $\lambda = 0.1/\pi $ and $\tau_{\rm R}=15$ for four-dimensional
		Schwarzschild black hole. The black contours $C_{i}$ are closed loops surrounding
		the zero points.}
\end{figure}

In this section, we investigate the topological number of the four-dimensional Schwarzschild black hole via the above topological approach. The Schwarzschild black hole metric is
\begin{equation}
	\label{schmetric}
	ds^{2}=-\left(1-\frac{2M}{r}\right)dt^{2}+\left(1-\frac{2M}{r}\right)^{-1}dr^{2}+r^{2}d\Omega_{2}^{2} ,
\end{equation}
\textcolor{red}{where $M$ is the ADM mass and $d\Omega^{2}_{2}$ is the line element of the unit $S^2$.} And the mass and entropy of the four-dimensional Schwarzschild black hole is given by
\begin{equation}
	\label{MSBH}
	M=\frac{r_{\rm h}}{2}, \quad S_{\rm BH}= \pi r_{\rm h}^{2},
\end{equation}
where $r_{\rm h}$ is the horizon radius. Using the transformation rule Eq.(\ref{transrule}), we obtain the R{\'e}nyi entropy function of Schwarzschild black hole as
\begin{equation}
	\label{SR}
	S_{\rm R}=\frac{1}{\lambda}\ln(1+\lambda \pi r_{\rm h}^2) .
\end{equation}
According to Eqs.(\ref{FR})(\ref{MSBH})(\ref{SR}). We obtain the generalized free energy as
\begin{equation}
	\mathcal{F} _{\rm R}=\frac{r_{\rm h}}{2}-\frac{1}{\lambda}\ln(1+\lambda \pi r_{\rm h}^2)\frac{1}{\tau_{\rm R}}.
\end{equation}
Then the components of the vector $\phi$ can be calculated as
\begin{equation}
	\begin{aligned}
		\phi ^{r_{\rm h}}&=\frac{1}{2}-\frac{2\pi r_{\rm h}}{1+\lambda \pi r_{\rm h}^{2}}\frac{1}{\tau_{\rm R}} , \\
		\phi ^{\Theta }&=-\cot \Theta   \csc \Theta  .
	\end{aligned}
\end{equation}
By solving the equation $\phi ^{r_{\rm h}}=0$, we obtain
\begin{equation}
	\label{tausch}
	\tau_{\rm R}=\frac{4\pi r_{\rm h}}{1+\lambda\pi r_{\rm h}^{2}}.
\end{equation}
Fig.\ref{fig.1-1} shows the curve of Eq.(\ref{tausch}) on the $r_{\rm h}-\tau_{\rm R}$ plane. When we fix the parameter $\lambda = 0.1/\pi $, we find that for $\tau_{\rm R} < \tau_{a}$ (e.g. $\tau_{\rm R}=\tau_{1}$), there are two points which satisfy the condition $\tau_{\rm R}=1/T_{\rm R}$, it reveals that there are two different types of black holes: one is thermodynamically stable, another is thermodynamically unstable, which are characterized by positive or negative values of the winding numbers.

In addition, we plot the unit vector field $n$ at $\tau_{\rm R}=15$ in Fig.\ref{fig.1-2}, where we find two zero points (ZP): ZP$_{1}$ at $r_{\rm h}=1.44$ and ZP$_{2}$ at $r_{\rm h}=6.94$, the corresponding winding numbers are $w_{1}=-1$ and $w_{2}=+1$, respectively. Thus, the topological number of four-dimensional Schwarzschild black hole is: $W=-1+1=0$.

Moreover, our calculation is based on the R{\'e}nyi statistics, which naturally leads to various result in~\cite{Wei:2022dzw} via the GB statistics. It is worth noting that previous study has shown that the canonical ensemble in flat spacetime which is described by the R{\'e}nyi formula exists just like in AdS spacetime~\cite{Czinner:2015eyk}. Now by comparing the topological number we calculated (including winding number) with the result of four-dimensional Schwarzschild-AdS black hole via the GB statistics, we find that they are similar, which indicates a connection  between the black hole thermodynamics in asymptotically flat spacetime via R{\'e}nyi statistics and that in asymptotically AdS spacetime via GB statistics. In the following section, we will analyze other types of black holes along with this thought.

%%%%%%%%%%%%%%%%%%%%%%%%%%%%%%%%%%%%%%%%%%%%%%%%%%%%%%%%%%%%%%%%%%%%%%
\section{The topological number of Reissner-Nordstr{\"o}m (RN) black holes}\label{RN topology}
%%%%%%%%%%%%%%%%%%%%%%%%%%%%%%%%%%%%%%%%%%%%%%%%%%%%%%%%%%%%%%%%%%%%%%

The topological number of RN black holes have been studied via the GB statistics in~\cite{Wei:2022dzw}, and it has been shown that the topological number of RN black hole is different from that of the Schwarzschild black hole. In the following, we will analyze the topological number for the charged case from the perspective of R{\'e}nyi statistics. We start with the $d$-dimensional RN black hole, the metric is
\begin{equation}	ds^{2}=-\left(1-\frac{2m}{r^{d-3}}+\frac{q^{2}}{r^{2(d-3)}}\right)dt^{2}+\left(1-\frac{2m}{r^{d-3}}+\frac{q^{2}}{r^{2(d-3)}}\right)^{-1}dr^{2}
	+r^{2}d\Omega_{d-2}^{2} ,
\end{equation}
where the ADM mass $M$, charge $Q$ and entropy $S_{\rm BH}$ of the black hole are
\begin{equation}
	\label{RNdthermodynamic quantities}
	M=\frac{d-2}{8\pi}\omega _{d-2}m,\quad Q=\frac{\sqrt{2(d-2)(d-3)} }{8\pi }\omega_{d-2} q ,\quad S_{\rm BH}=\frac{\omega _{d-2}r_{\rm h}^{d-2}}{4} ,
\end{equation}
where $r_{\rm h}$ is the outer horizon radius and $\omega_{d-2}=2\pi^{(d-1)/2}/ \Gamma ((d-1)/2)$ is the volume of the unit $S^{d-2}$. In following subsection, we will discuss the four-dimensional and higher dimensional cases based on the above thermodynamic quantities.
	\subsection{Four-dimensional  case}
For the case $d=4$, $\omega_{2}=4 \pi $, the thermodynamic quantities (\ref{RNdthermodynamic quantities}) reduce to
\begin{equation}
	\label{RNmass}
	M=m, \quad Q=q, \quad S_{\rm BH}=\pi r_{\rm h}^{2}.
\end{equation}
Thus, the R{\'e}nyi entropy is given by
\begin{equation}
	\label{RN4renyi}
	S_{\rm R}=\frac{1}{\lambda} \ln (1+\lambda \pi r_{\rm h}^{2}).
\end{equation}
From Eqs.(\ref{RNmass})(\ref{RN4renyi}), we obtain the generalized free energy
\begin{equation}
	\begin{aligned}
		\mathcal{F}_{\rm R}&=M-\frac{S_{\rm R}}{\tau_{\rm R}}\\
		&=\frac{r_{\rm h}}{2}+\frac{Q^{2}}{2r_{\rm h}}-\frac{1}{\lambda}\ln(1+\lambda \pi r_{\rm h}^2)\frac{1}{\tau_{\rm R}}.
	\end{aligned}
\end{equation}
The components of the vector $\phi$ can be calculated as
\begin{equation}
	\begin{aligned}
		\phi ^{r_{\rm h}}& = \frac{1}{2}-\frac{Q^{2}}{2r_{\rm h}^{2}} -\frac{2\pi r_{\rm h}}{1+\lambda \pi r_{\rm h}^{2}}\frac{1}{\tau_{\rm R}},\\
		\phi ^{\Theta }&=-\cot \Theta \csc \Theta,
	\end{aligned}
\end{equation}
\begin{figure}[htbp]
	\centering
	\subfloat[$\lambda <\lambda_{c} $]{\includegraphics[width=0.45\columnwidth,height=0.45\linewidth]{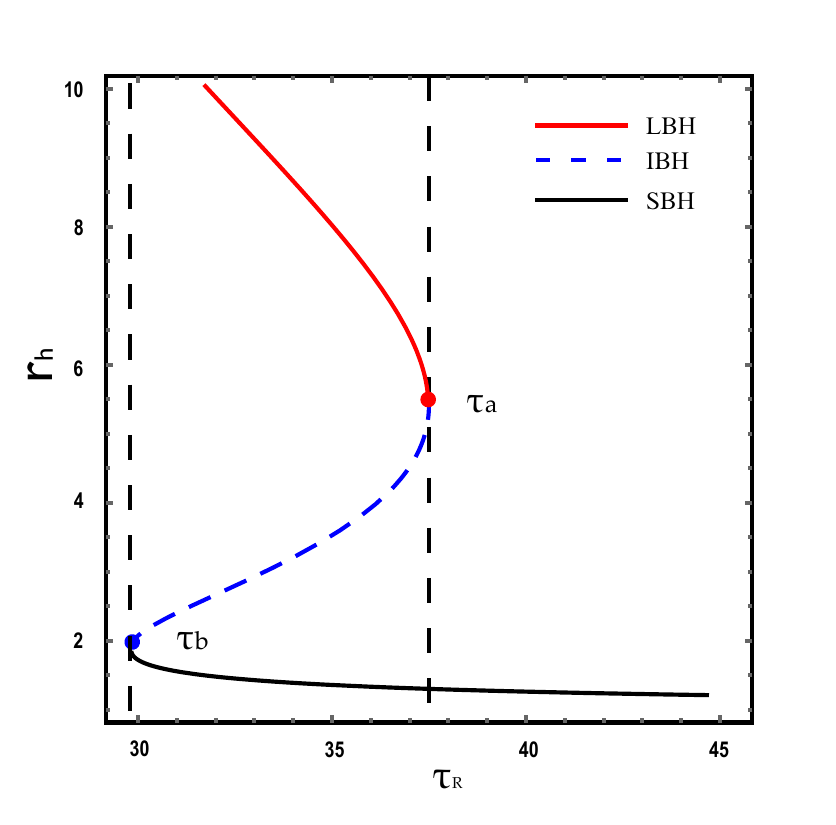}}
	\quad
	\subfloat[$\lambda > \lambda_{c} $]{\includegraphics[width=0.45\columnwidth,height=0.45\linewidth]{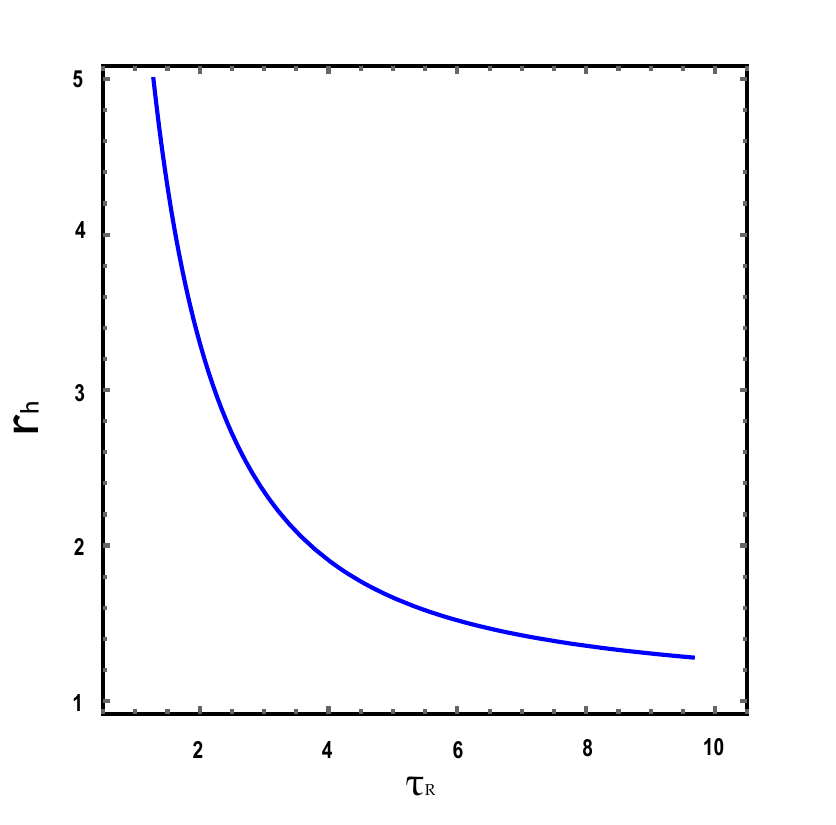}}
	
	\caption{The curves of equation (\ref{tuaRRN}), where the stable and unstable black hole
		branches are plotted in solid and dashed lines, respectively. The left figure (a) is plotted with $Q=1$, $\lambda=0.03/\pi <\lambda_{c}$ and $\tau_{\rm R}=36$. The red solid, blue dashed, and black solid lines are for the large black hole (LBH), intermediate black hole (IBH), and small black hole (SBH), respectively. The annihilation and generation points are represented by red and blue dots, respectively. The right figure (b) is plotted with $Q=1$, $\lambda=2/\pi  >\lambda_{c}$ and $\tau_{\rm R}=1.5$.}
	\label{fig.2}
\end{figure}
and the on-shell condition $\phi^{r_{\rm h}}=0$ gives
\begin{equation}
	\label{tuaRRN}
	\tau_{\rm R}=\frac{4\pi r_{\rm h}^{3}}{(r_{\rm h}^{2}-Q^{2})(1+\lambda \pi r_{\rm h}^{2} )} .
\end{equation}

\begin{figure}[htbp]
	\centering
	\subfloat[$\lambda <\lambda_{c} $]{\includegraphics[width=0.48\columnwidth,height=0.48\linewidth]{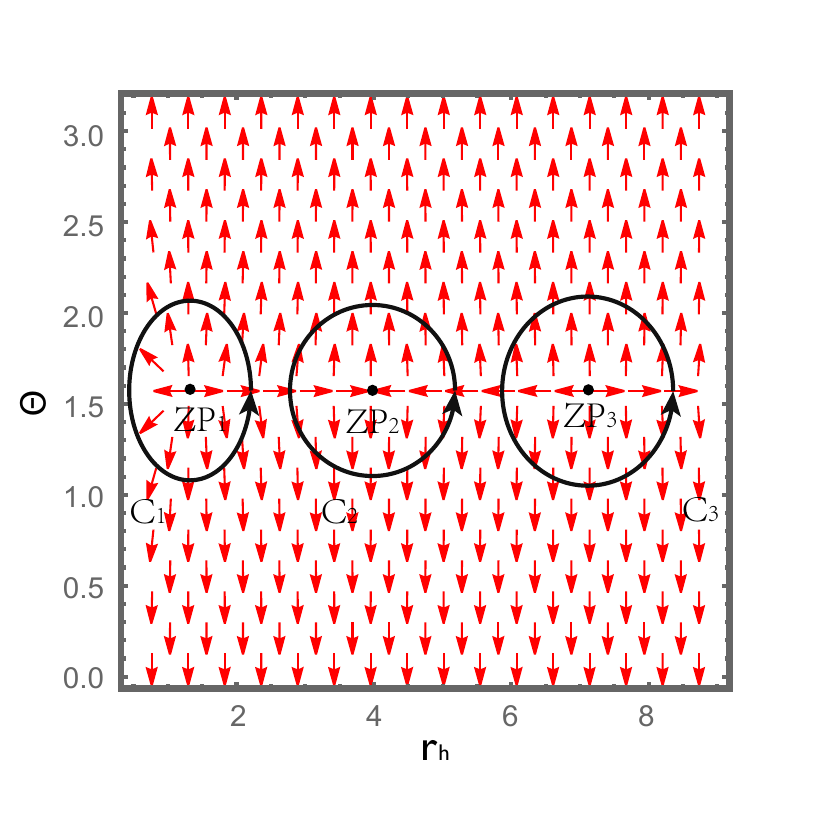}\label{fig.2vector1}}
	\quad
	\subfloat[$\lambda > \lambda_{c} $]{\includegraphics[width=0.48\columnwidth,height=0.48\linewidth]{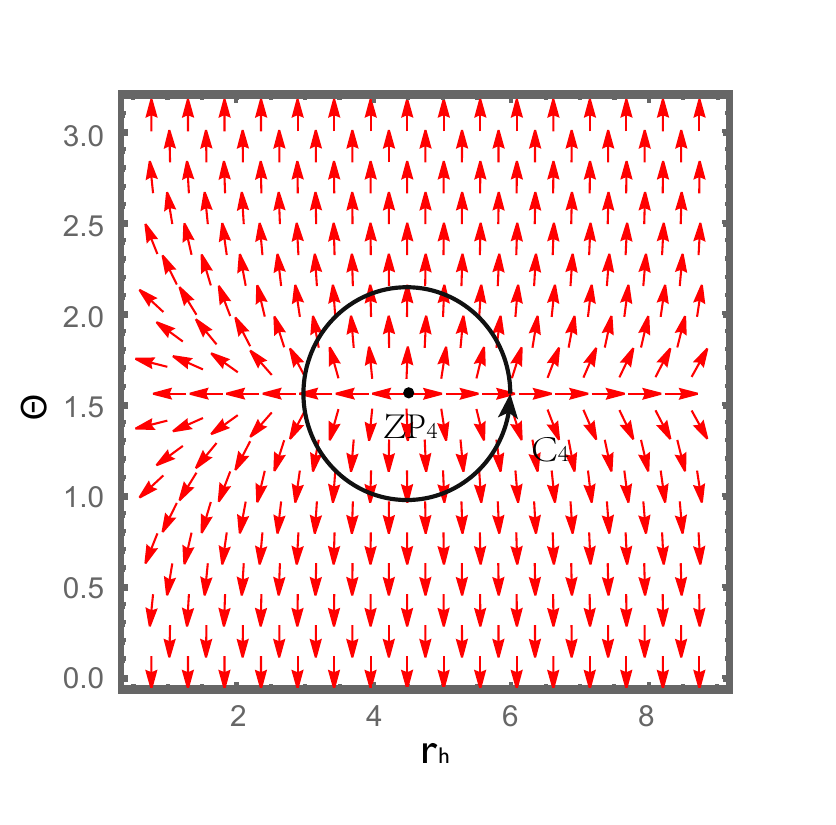}\label{fig.2vector2}}
	\caption{The red arrows represent the unit vector field $n$ on a portion of the $r_{\rm h}-\Theta$ plane for four-dimensional RN black hole. The left figure (a) is plotted with $Q=1$ and $\lambda=0.03/\pi <\lambda_{c}$. The right figure (b) is plotted with $Q=1$, $\lambda=2/\pi >\lambda_{c}$.}
	\label{fig.2vector}
\end{figure}
Next, we will give the curve between $r_{\rm h}$ and $\tau_{\rm R}$ when the parameter $Q$ and $\lambda$ are fixed. Noting that there is a critical parameter $\lambda_{c}=(7-4\sqrt{3})/\pi Q^{2}$ for the four-dimensional black hole, when $\lambda<\lambda_{c}$, there is a SBH/LBH phase transition, otherwise, i.e., $\lambda>\lambda_{c}$, there is no phase transition~\cite{ Promsiri:2020jga}. From Fig.\ref{fig.2}, when choosing $Q=1$, $\lambda=0.03/\pi <\lambda_{c}$, we can find that there are three black hole branches: the red and black solid lines for large and small black hole branches, respectively, and the blue dashed line for intermediate black hole branch. When taking $Q=1$, $\lambda=2/\pi  >\lambda_{c}$, there is only one black hole branch.

Besides, the unit vector field is plotted in Fig.\ref{fig.2vector1}, in which there are three zero points: ZP$_{1}$ at $r_{\rm h}=1.34$,  ZP$_{2}$ at $r_{\rm h}=3.89$ and ZP$_{3}$ at $r_{\rm h}=7.28$. The winding numbers of both large black hole branch (e.g. ZP$_{3}$) and small black hole branch (e.g. ZP$_{1}$) are equal to $+1$, while the winding number of intermediate black hole branch (e.g. ZP$_{2}$) is equal to $-1$. Thus, the topological number of four-dimensional RN black hole is: $W=+1-1+1=+1$. For the case $\lambda>\lambda_{c}$, only one zero point ZP$_{4}$ at $r_{\rm h}=4.31$ As shown in Fig.\ref{fig.2vector2}, which has the same topological number as the case $\lambda<\lambda_{c}$, i.e. $W=+1$.

\subsection{Higher dimensional cases}
\label{RNhigherd}
For the case of higher dimensional RN black hole, it is convenient to study its topological numbers by analyzing the asymptotic behaviors of $\tau_{\rm R}(r_{\rm h})$ at small and large $r_{\rm h}$ limits. Starting with thermodynamic quantities in Eq.(\ref{RNdthermodynamic quantities}), the generalized free energy is given by
\begin{equation}
	\begin{aligned}
		\mathcal{F}_{\rm R} & =M-\frac{S_{\rm R}}{\tau_{\rm R}}\\
		&= \frac{d-2}{8\pi }\omega _{d-2}(\frac{r_{\rm h}^{d-3}}{2}+\frac{q^{2}}{2r_{\rm h}^{d-3}}  )-\frac{\ln (1+\lambda\frac{\omega _{d-2}r_{\rm h}^{d-2}}{4}  )}{\lambda \tau_{\rm R}}.
	\end{aligned}
\end{equation}
Thus, the zero point $\partial \mathcal{F}_{\rm R}/\partial r_{\rm h}=0$ is
\begin{equation}
	\label{tauRNd}
	\tau_{\rm R}=\frac{16\pi r_{\rm h}^{2d-5}}{(d-3)(r_{\rm h}^{2d-6}-q^{2})(4+\lambda r_{\rm h}^{d-2}\omega _{d-2})} .
\end{equation}
When the charge $q=0$, Eq.(\ref{tauRNd}) reduces to
\begin{equation}
	\label{tauRsch}
	\tau_{\rm R}=\frac{16\pi r_{\rm h}}{(d-3)(4+\lambda r_{\rm h}^{d-2}\omega _{d-2})} ,
\end{equation}
which is the case of $d$-dimensional Schwarzschild black hole, the lower bound of $r_{\rm h}$ in Eq.(\ref{tauRsch}) is zero, i.e. $r_{\rm min}=0$, so the asymptotic of $\tau_{\rm R}$ in small and large $r_{\rm h}$ limits satisfy
\begin{equation}
	\begin{aligned}
		&\tau_{\rm R}\to 0, \quad r_{\rm h}\to \infty ,\\
		&\tau_{\rm R}\to 0, \quad r_{\rm h}\to 0 ,
	\end{aligned}
\end{equation}
which is similar to Type 4 introduced in \cite{Liu:2022aqt}, and the topological number is equal to $0$.

When the charge $q$ is nonzero, the lower bound of $r_{\rm h}$ in Eq.(\ref{tauRNd}) is the horizon radius of the extremal black hole with zero temperature (i.e. $r_{\rm min}=r_{\rm ex}=q^{\frac{1}{d-3}}$), then we obtain
\begin{equation}
	\begin{aligned}
		&\tau_{\rm R}\to 0, \quad r_{\rm h}\to \infty ,\\
		&\tau_{\rm R}\to \infty, \quad r_{\rm h}\to r_{\rm ex} ,
	\end{aligned}
\end{equation}
which is similar to Type 2 introduced in \cite{Liu:2022aqt}, and the topological number $W=+1$. From the above discussion, we can see that the charge has an effect on the topological number of the black hole, and that the dimension has no effect on the topological number of  Schwarzschild and RN black holes calculated via R{\'e}nyi statistics, which is similar to the result by using the GB statistics in \cite{Wei:2022dzw}.
%%%%%%%%%%%%%%%%%%%%%%%%%%%%%%%%%%%%%%%%%%%%%%%%%%%%%%%%%%%%%%%%%%%%%%
\subsubitem
\section{The topological number of Kerr black holes}\label{Kerr topology}
%%%%%%%%%%%%%%%%%%%%%%%%%%%%%%%%%%%%%%%%%%%%%%%%%%%%%%%%%%%%%%%%%%%%%%

In this section, we will explore the cases for rotating black holes, i.e., Kerr black hole via the R{\'e}nyi statistics. For $d$-dimensional singly rotating Kerr black hole, its metric is
\begin{equation}
	\label{kerrmetric}
	\begin{aligned}
		d s^{2}= & -\frac{\Delta_{r}}{\Sigma}\left(d t-a \sin ^{2} \theta d \phi\right)^{2}+\frac{\Sigma}{\Delta_{r}} d r^{2}+\Sigma d \theta^{2} \\
		& +\frac{\sin ^{2} \theta}{\Sigma}\left[a d t-\left(r^{2}+a^{2}\right) d \phi\right]^{2}+r^{2} \cos ^{2} \theta d \Omega_{d-4}^{2},
	\end{aligned}
\end{equation}
where
\begin{equation}
	\Delta _{r}=r^{2}+a^{2}-2mr^{5-d},\quad \quad \Sigma =r^{2}+a^{2}\cos \theta,
\end{equation}
and its corresponding mass, horizon entropy, angular velocity and angular momentum are respectively
\begin{equation}
	\label{thermodynamic quantities}
	\begin{aligned}
		&M=\frac{d-2}{8\pi}\omega _{d-2}m,\quad S_{\rm BH}=\frac{\omega _{d-2}}{4}(r_{\rm h}^{2}+a^{2})r_{\rm h}^{d-4}, \\
		&\Omega=\frac{a}{r_{\rm h}^{2}+a^{2}},\quad J=\frac{\omega _{d-2}}{4\pi }ma,
	\end{aligned}
\end{equation}
where the horizon radius $r_{\rm h}$ is given by the equation $\Delta _{r}=0$.
\subsection{Four-dimensional case}
We start with the case $d=4$, the thermodynamic quantities in Eq.(\ref{thermodynamic quantities}) reduces to
\begin{equation}
	\begin{aligned}
		&M=m,\quad S_{\rm BH}=\pi(r_{\rm h}^{2}+a^{2}), \\
		&\Omega=\frac{a}{r_{\rm h}^{2}+a^{2}},\quad J=ma.
	\end{aligned}
\end{equation}
Therefore, the R{\'e}nyi entropy is calculated as follows
\begin{equation}
	S_{\rm R}=\frac{1}{\lambda}\ln(1+\lambda \pi (r_{\rm h}^2+a^{2})) ,
\end{equation}
and the generalized free energy of the four-dimensional Kerr black hole can be written as
\begin{equation}
	\begin{aligned}
		\mathcal{F}_{\rm R}&=M-\frac{S_{\rm R}}{\tau_{\rm R}}\\
		&=\frac{r_{\rm h}^{2}+a^{2}}{2 r_{\rm h}}-\frac{1}{\lambda}\ln(1+\lambda \pi (r_{\rm h}^2+a^{2}))\frac{1}{\tau_{\rm R}}	.
	\end{aligned}
\end{equation}
\begin{figure}[htbp]
	\centering
	\subfloat[$\lambda <\lambda_{c} $]{\includegraphics[width=0.45\columnwidth,height=0.45\linewidth]{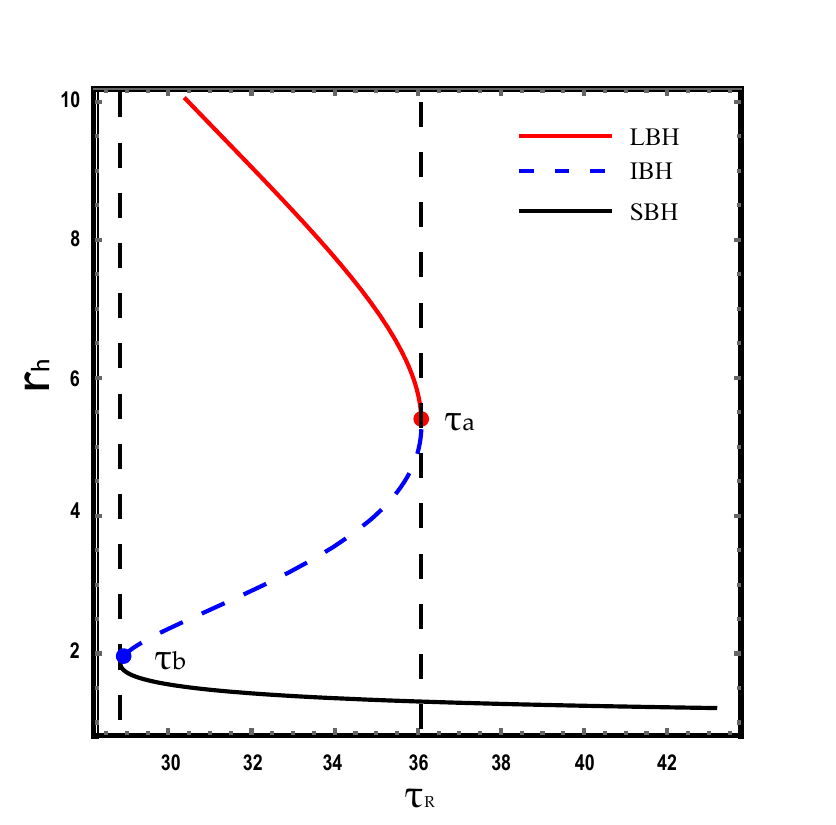}}
	\quad
	\subfloat[$\lambda > \lambda_{c} $]{\includegraphics[width=0.45\columnwidth,height=0.45\linewidth]{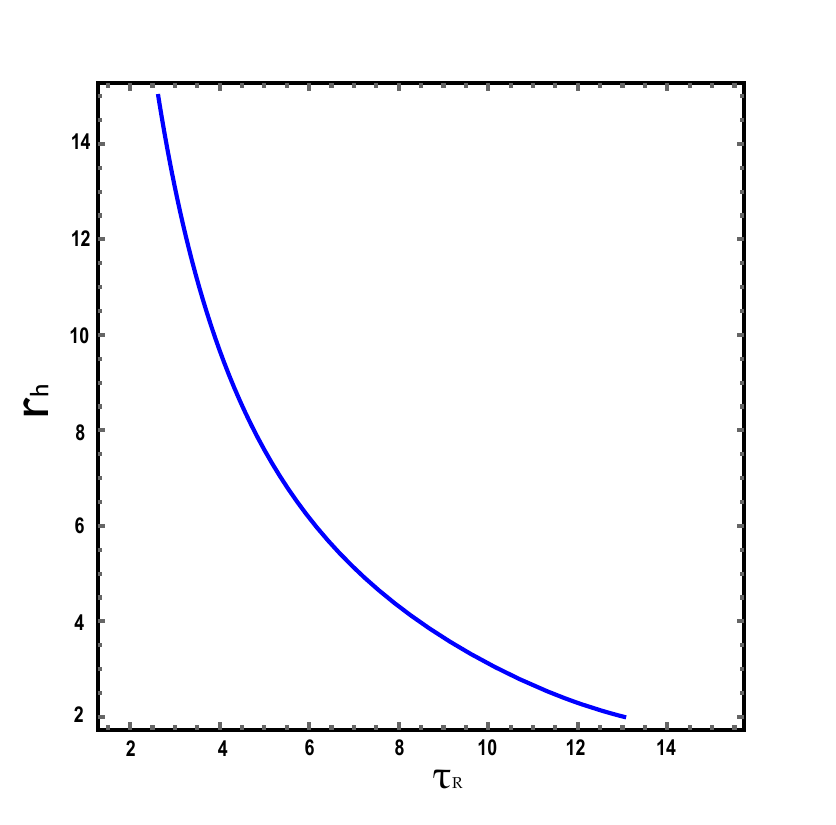}}
	
	\caption{The curves of Eq.(\ref{tauRKerr}), where the stable and unstable black hole
		branches are plotted in solid and dashed lines, respectively. The left figure (a) is plotted with $a=1$ and $\lambda=0.01  <\lambda_{c}$. The red solid, blue dashed, and black solid lines are for the large black hole (LBH), intermediate black hole (IBH), and small black hole (SBH), respectively. The annihilation and generation points are represented by red and blue dots, respectively. The right figure (b) is plotted with $a=1$, $\lambda=0.1 >\lambda_{c}$.}
	\label{fig.3}
\end{figure}
The components of the vector $\phi$ can be given by
\begin{equation}
	\begin{aligned}
		\phi ^{r_{\rm h}}& = \frac{1}{2}-\frac{a^2}{2r_{\rm h}^{2}} -\frac{2\pi r_{\rm h}}{1+\lambda \pi (r_{\rm h}^{2}+a^{2})}\frac{1}{\tau_{\rm R}},\\
		\phi ^{\Theta }&=-\cot \Theta \csc \Theta .
	\end{aligned}
\end{equation}
The zero points are determined by $\phi=0$, namely,
\begin{equation}
	\label{tauRKerr}
	\tau_{\rm R}=\frac{4\pi r_{\rm h}^{3}}{(r_{\rm h}^{2}-a^{2})(1+\lambda \pi (r_{\rm h}^{2}+a^{2}) )} .
\end{equation}
\begin{figure}[htbp]
	\centering
	\subfloat[$\lambda <\lambda_{c} $]{\includegraphics[width=0.45\columnwidth,height=0.45\linewidth]{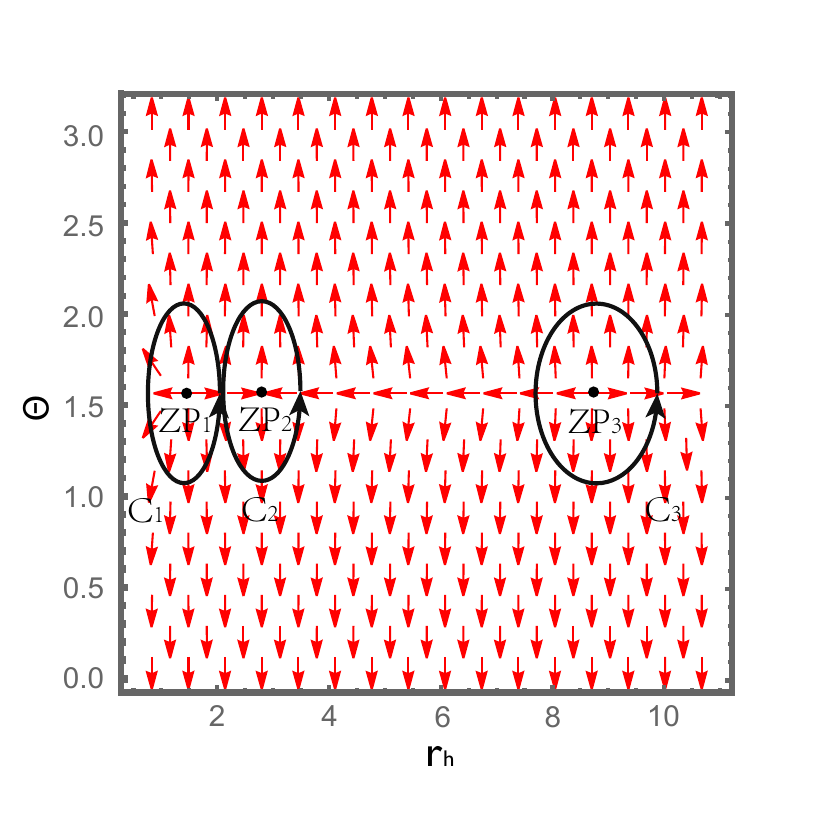}\label{fig.3vector1}}
	\quad
	\subfloat[$\lambda > \lambda_{c} $]{\includegraphics[width=0.45\columnwidth,height=0.45\linewidth]{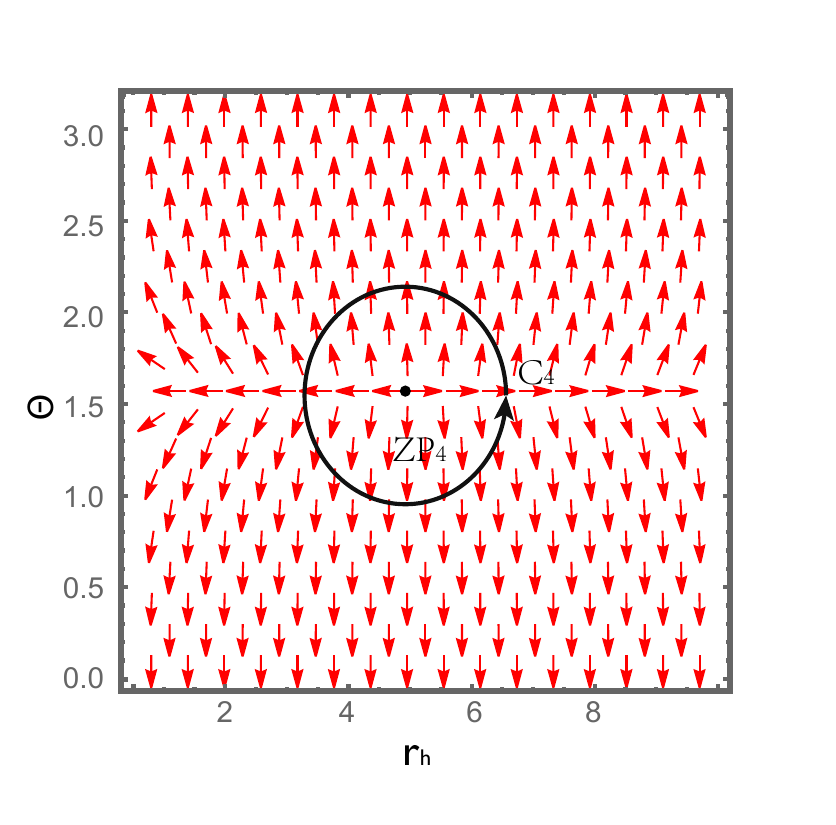}\label{fig.3vector2}}
	\caption{The red arrows represent the unit vector field $n$ in a portion of the $r_{\rm h}-\Theta$ plane for four-dimensional Kerr black hole. The left figure (a) is plotted with $a=1$, $\lambda=0.01 <\lambda_{c}$ and $\tau_{\rm R} =32$. The right figure (b) is plotted with $a=1$, $\lambda=0.1 >\lambda_{c}$ and $\tau_{\rm R} =7$.}
	\label{fig.3vector}
\end{figure}

Through the analysis of Eq.(\ref{tauRKerr}), we find that there is a critical parameter $\lambda_{c}$ for the four-dimensional rotating case. It is similar to the case of four-dimensional RN black hole, where there is a phase transition or no phase transition when the value of parameter $\lambda$ is different (i.e., $\lambda<\lambda_{c}$ or $\lambda>\lambda_{c}$). For more discussion on the phase transition, refer to \cite{Czinner:2017tjq}. To plot the curve of Eq.(\ref{kerrmetric}), we set $a=1$, and choose various values for  $\lambda =0.01<\lambda_{c}$ and $\lambda=0.1>\lambda_{c}$. As shown in Fig. \ref{fig.3}, there are two types of curves on the $r_{\rm h}-\tau_{\rm R}$ plane, which is the same as that of the analysis of four-dimensional RN black hole. In Fig.\ref{fig.3vector}, we plot the vector field, there are three ZPs in Fig.\ref{fig.3vector1}: ZP$_{1}$ at $r_{\rm h}=1.43$,  ZP$_{2}$ at $r_{\rm h}=2.92$ and ZP$_{3}$ at $r_{\rm h}=9.01$, only one zero point in Fig.\ref{fig.3vector2}: ZP$_{4}$ at $r_{\rm h}=5.12$. Thus, one can make a similar calculation of the topological number, and obtain the topological number $W$ for each cases $\lambda<\lambda_{c}$ and $\lambda>\lambda_{c}$ is equal to $+1$.

\subsection{Higher dimensional cases}
For the $d>4$-dimensional Kerr black hole, from Eq.(\ref{thermodynamic quantities}), its R{\'e}nyi entropy and generalized free energy are
\begin{equation}
	\begin{aligned}
		S_{\rm R} & = \frac{1}{\lambda }\ln (1+\frac{\omega _{d-2}}{4} \lambda(r_{\rm h}^{2}+a^{2})r_{\rm h}^{d-4} ),\\
		\mathcal{F}_{\rm R}  &= M-\frac{S_{\rm R}}{\tau_{\rm R}}\\
		&=\frac{(d-2)\omega _{d-2}(r_{\rm h}^{2}+a^{2})}{16\pi r_{\rm h}^{5-d}} -\frac{1}{\lambda }\frac{\ln (1+\frac{\omega _{d-2}}{4} \lambda(r_{\rm h}^{2}+a^{2})r_{\rm h}^{d-4} )}{\tau _{\rm R}}.
	\end{aligned}
\end{equation}
Then the components of the vector $\phi$ can be given by
\begin{equation}
	\begin{aligned}
		\phi ^{r_{\rm h}}  = &\frac{(d-2)r_{\rm h}^{d-4}\omega _{d-2}}{8\pi }+\frac{(d-2)(d-5)r_{\rm h}^{d-6}(a^{2}+r_{\rm h}^{2})w_{d-2}}{16\pi }\\
		&-\frac{(2r_{\rm h}^{d-3}+(d-4)r_{\rm h}^{d-5}(a^{2}+r_{\rm h}^{2})) \omega _{d-2}}{\tau_{\rm R}(4+r_{\rm h}^{d-4}(a^{2}+r_{\rm h}^{2})\lambda \omega _{d-2})}\\
		\phi ^{\Theta}=& -\cot \Theta \csc \Theta .
	\end{aligned}
\end{equation}
\begin{figure}[htbp]
	\centering
	\subfloat[]{\includegraphics[width=0.46\columnwidth,height=0.46\linewidth]{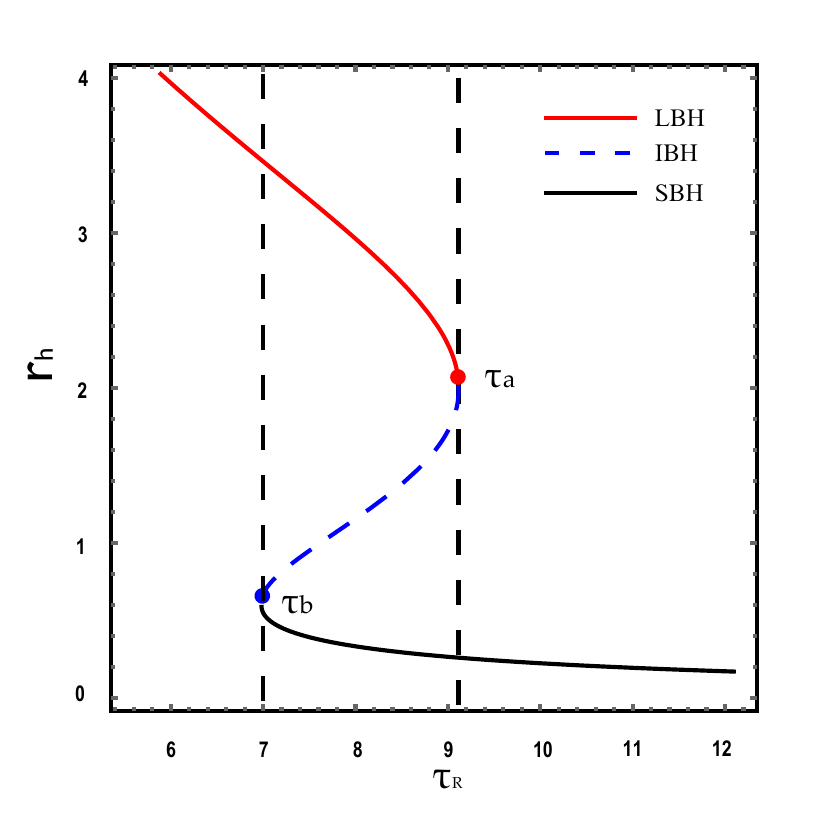}\label{fig.3-5}}
	\quad
	\subfloat[]{\includegraphics[width=0.48\columnwidth,height=0.48\linewidth]{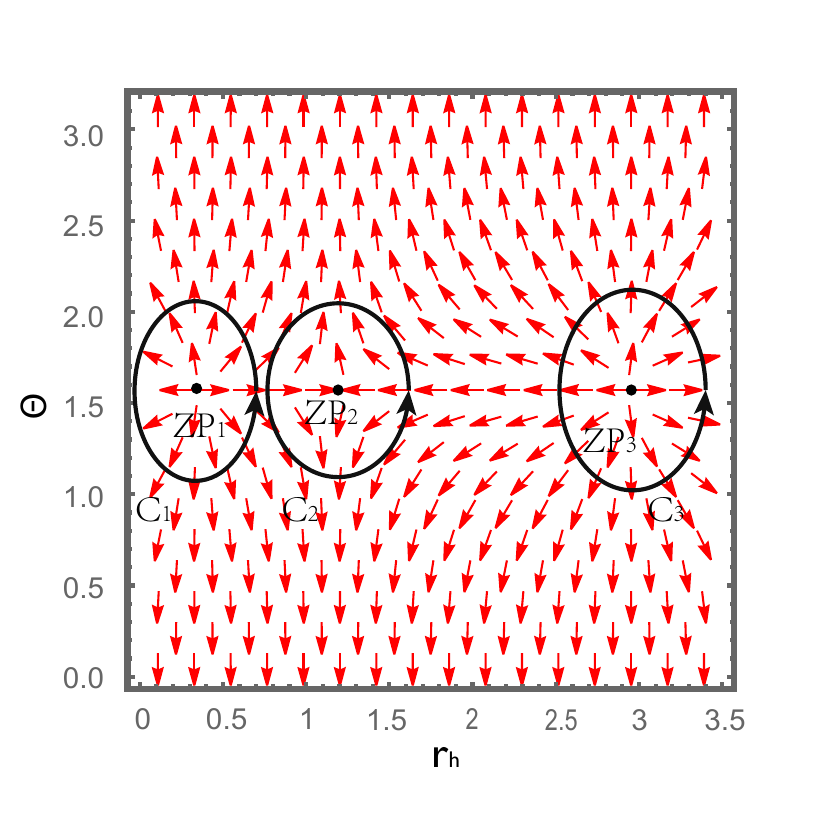}\label{fig.3-6}}
	
	\caption{The left figure (a): zero points of the vector $\phi^{r_{\rm h}}$ shown in the $r_{\rm h}-\tau_{\rm R}$ plane with $a=1$ and $\lambda= 0.01 $ for five-dimensional singly rotating Kerr black hole. The red solid, blue dashed, and black solid lines are for the large black hole (LBH), intermediate black hole (IBH), and small black hole (SBH), respectively. The annihilation and generation points are represented by red and blue dots, respectively. The left figure (b): the red arrows represent the unit vector field $n$ on a portion of the $r_{\rm h}-\Theta$ plane with $\lambda= 0.01 $, $a=1$ and $\tau_{\rm R}=8$ for the case $d=5$. The black contours $C_{i}$ are closed loops surrounding
		the zero points.}
\end{figure}
By solving $\phi^{r_{\rm h}}=\partial{\mathcal{F}_{\rm R} }/\partial {r_{\rm h}}=0 $, the relationship between $\tau_{\rm R}$ and $r_{\rm h}$ is given by
\begin{equation}
	\label{kerrdtau}
	\tau _{\rm R}=\frac{16\pi r_{\rm h}^{5}((d-4)a^{2}+(d-2)r_{\rm h}^{2})}{(d-2)((d-3)r_{\rm h}^{2}+(d-5)a^{2})(4r_{\rm h}^{4}+\lambda  \omega _{d-2}r_{\rm h}^{d}(a^{2}+r_{\rm h}^{2}))} .
\end{equation}
Next, we will firstly discuss situations in specific dimensions, and finally give an analysis from the asymptotic behaviors of $\tau_{\rm R}$.

{
	\centering
	\subsubsection*{$(a)$.\ $d=5$}
}

When $d=5$, $\omega_{3}=2 \pi^{2} $, the vector $\phi$ is given by
\begin{equation}
	\begin{aligned}
		\phi ^{r_{\rm h}}& = \frac{3 \pi r _{\rm h}}{4}-\frac{\pi^{2}\left(a^{2}+3 r_{\rm h}^{2}\right)}{\left(2+a^{2} \pi^{2} r_{\rm h} \lambda+\pi^{2} r_{\rm h}^{3} \lambda\right) \tau_{\rm R}},\\
		\phi ^{\Theta }&=-\cot \Theta \csc \Theta ,
	\end{aligned}
\end{equation}
and the zero point of vector satisfies
\begin{equation}
	\label{kerr5tau}
	\tau_{\rm R}=\frac{4\pi (a^{2}+3r_{\rm h}^{2})}{3r_{\rm h}(2+\lambda  \pi^{2} r_{\rm h}(a^{2}+r_{\rm h}^{2}))}.
\end{equation}
In Fig.\ref{fig.3-5}, we plot the zero point of $\phi^{r_{\rm h}}$ by fixing $a=1$, $\lambda=0.01$, and there are three black hole branches. The unit vector field $n$ with $\tau_{\rm R}=8$ is shown in Fig.\ref{fig.3-6}, there are three zero points which are located at ZP$_i$ $(r_{\rm h},\Theta)=(0.35,\pi/2), \ (1.15,\pi/2), \ (2.94,\pi/2)$. From Figs.\ref{fig.3-5} and \ref{fig.3-6}, the topological number of five-dimensional rotating black hole can be calculated as: $W=+1-1+1=+1$, where $+1$ is the winding number of large and small black hole branches, and $-1$ is the winding number of intermediate black hole branch. Furthermore, we can find that the topological number of five-dimensional case is the same as that of the four-dimensional case.

{
	\centering
	\subsubsection*{$(b)$.\ $d=6$}
}
\begin{figure}[htbp]
	\centering
	\subfloat[]{\includegraphics[width=0.46\columnwidth,height=0.46\linewidth]{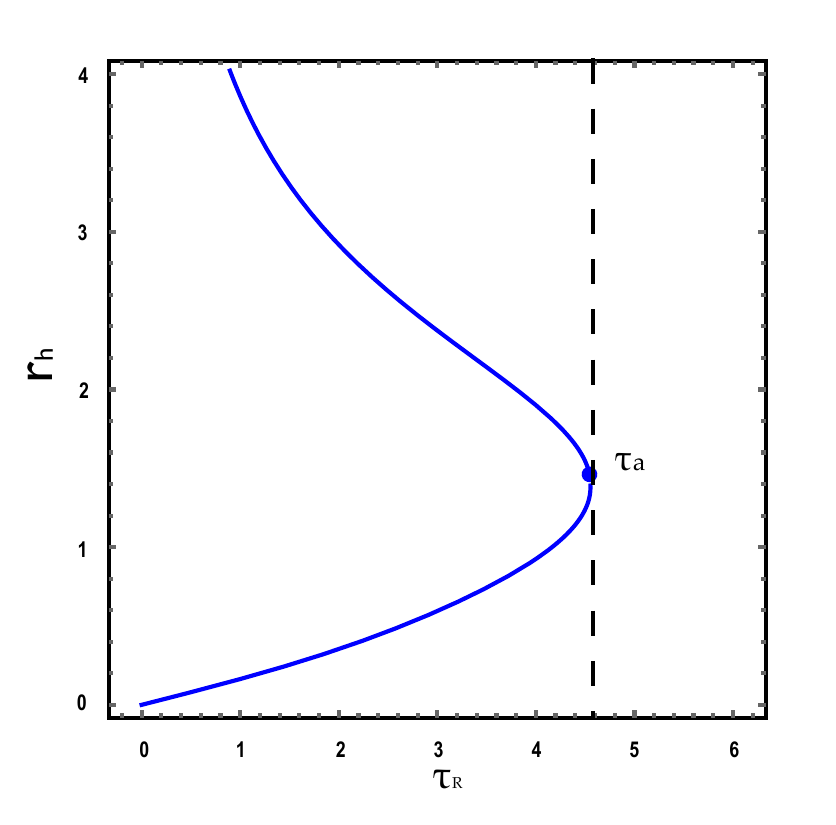}\label{fig.3-7}}
	\quad
	\subfloat[]{\includegraphics[width=0.48\columnwidth,height=0.48\linewidth]{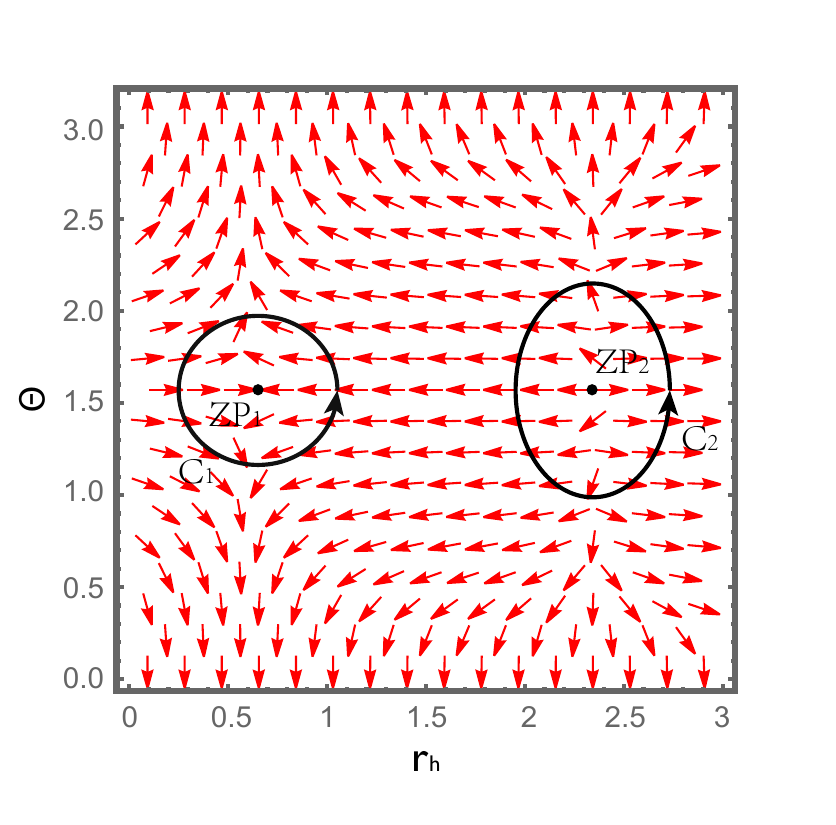}\label{fig.3-8}}
	
	\caption{The left figure (a): zero points of the vector $\phi^{r_{\rm h}}$ shown in the $r_{\rm h}-\tau_{\rm R}$ plane with $a=1$ and $\lambda= 0.01 $ for six-dimensional singly rotating Kerr black hole. The annihilation point for this black hole is represented by the blue dot with $\tau_{a}$. The left figure (b): the red arrows represent the unit vector field $n$ on a portion of the $r_{\rm h}-\Theta$ plane with $\lambda= 0.01 $, $a=1$ and $\tau_{\rm R}=3$ for the
		case $d=6$. The black contours $C_{i}$ are closed loops surrounding
		the zero points.}
\end{figure}
For the $d=6$ case, $\omega_{4}=8\pi^{2}/3$, the components of vector $\phi$ are given by
\begin{equation}
	\begin{aligned}
		\phi ^{r_{\rm h}}&=\frac{4\pi r_{\rm h}^{2}}{3}+\frac{2\pi }{3}(a^{2}+r_{\rm h}^{2})-\frac{4\pi ^{2}r_{\rm h}(a^{2}+2r_{\rm h}^{2})}{(3+2\pi^{2} \lambda r_{\rm h}^{2}(a^{2}+r_{\rm h}^{2}) )\tau_{\rm R}}    \\
		\phi ^{\Theta }&=-\cot \Theta \csc \Theta,
	\end{aligned}
\end{equation}
and the zero point of the vector gives
\begin{equation}
	\label{kerr6tau}
	\tau_{\rm R}=\frac{6\pi r_{\rm h}(a^{2}+2r_{\rm h}^{2})}{(a^{2}+3r_{\rm h}^{2})(3+2\pi^{2}r_{\rm h}^{2}\lambda (a^{2}+r_{\rm h}^{2}) )} .
\end{equation}
Taking the parameter $a=1$ and $\lambda=0.01$, we plot the curve of Eq.(\ref{kerr6tau}) in Fig.\ref{fig.3-7}, and the unit vector $n$ with $\tau_{\rm R}=3$ is shown in Fig.\ref{fig.3-8}, one can easily calculate the winding numbers of two points which are located at ZP$_{1}$ $(0.59,\pi/2)$ and ZP$_{2}$ $(2.35,\pi/2)$ as: $\omega_{1}=-1$, $\omega_{2}=+1$, respectively. Thus, the topological number $W$ of six-dimensional case is equal to $-1+1=0$, which different from the topological number for the cases $d=4,5$.

{
	\centering
	\subsubsection*{$(c)$.\ $d=7$}
}

\begin{figure}[htbp]
	\centering
	\subfloat[]{\includegraphics[width=0.46\columnwidth,height=0.46\linewidth]{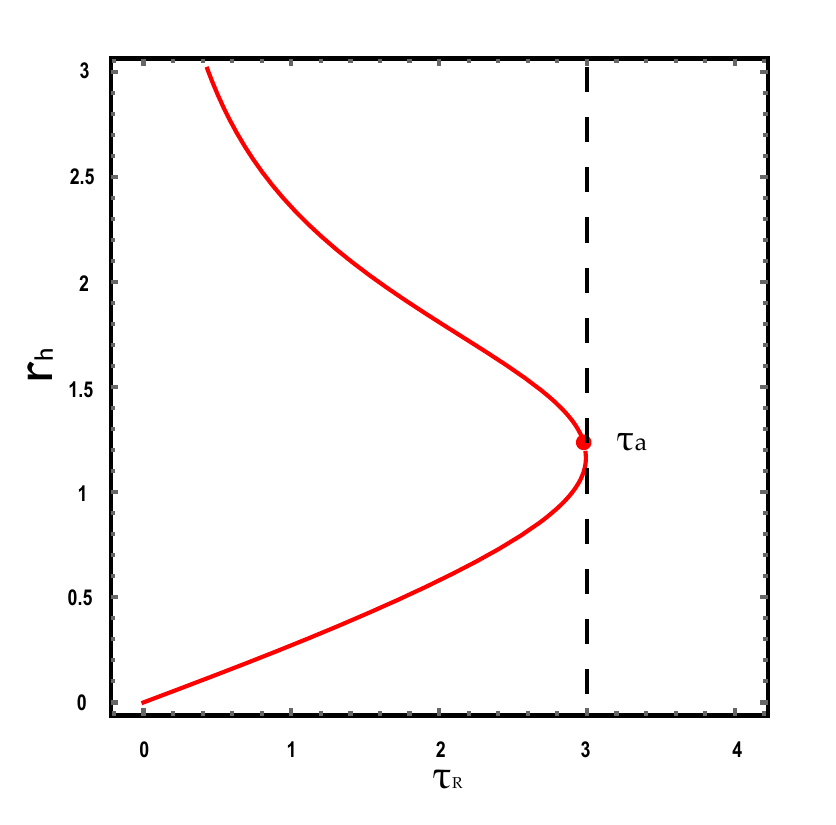}\label{fig.3-9}}
	\quad
	\subfloat[]{\includegraphics[width=0.48\columnwidth,height=0.48\linewidth]{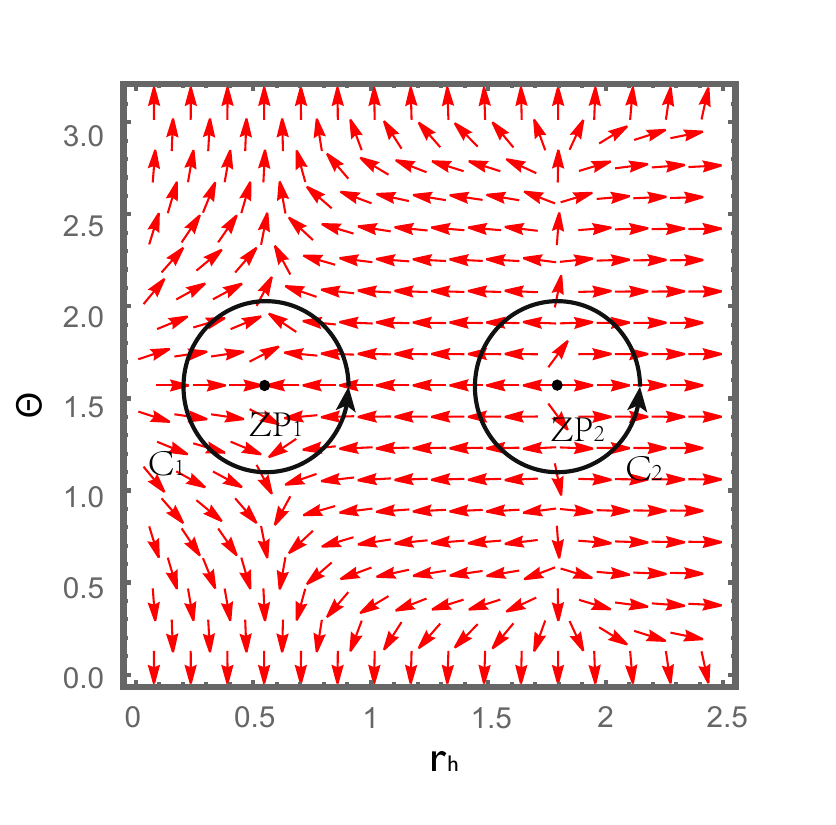}\label{fig.3-10}}
	
	\caption{The left figure (a): zero points of the vector $\phi^{r_{\rm h}}$ shown in the $r_{\rm h}-\tau_{\rm R}$ plane with $a=1$ and $\lambda= 0.01 $ for seven-dimensional singly rotating Kerr black hole. The annihilation point for this black hole is represented by the red dot with $\tau_{a}$. The left figure (b): the red arrows represent the unit vector field $n$ on a portion of the $r_{\rm h}-\Theta$ plane with $\lambda = 0.01 $, $a=1$ and $\tau_{\rm R}=2$ for the
		case $d=7$. The black contours $C_{i}$ are closed loops surrounding
		the zero points.}
\end{figure}

As for the $d=7$ case, and the volume of unit $S^7$ is $\omega_{5}=\pi^{3}$ and the vector $\phi^{a}$ is given by
\begin{equation}
	\begin{aligned}
		\phi ^{r_{\rm h}}&=\frac{5\pi^{2}r_{\rm h}^{3} }{8}+\frac{5\pi ^{2}}{8}r_{\rm h}(a^{2}+r_{\rm h}^{2})-\frac{\pi ^{3}r_{\rm h}^{2}(3a^{2}+5r_{\rm h}^{2})}{(4+\lambda \pi r_{\rm h}^{3}(a^{2}+r_{\rm h}^{2}))\tau_{\rm R}}   \\
		\phi ^{\Theta }&=-\cot \Theta \csc \Theta,
	\end{aligned}
\end{equation}
and thezero point of the vector gives
\begin{equation}
	\label{kerr7tau}
	\tau_{\rm R}=\frac{8\pi r_{\rm h}(3a^{2}+5r_{\rm h}^{2})}{5(a^{2}+2r_{\rm h}^{2})(4+\pi^{3}r_{\rm h}^{3}\lambda (a^{2}+r_{\rm h}^{2}) )} .
\end{equation}
Fig.\ref{fig.3-9} is the curve of Eq.(\ref{kerr7tau}) with $a=1$ and $\lambda=0.01$, and Fig.\ref{fig.3-10} is the unit vector $n$ with $\tau_{\rm R}=2$, which exists two points ZP$_{1}$ $(0.58,\pi/2)$ and ZP$_{2}$ $(1.80,\pi/2)$. From Figs.\ref{fig.3-9} and \ref{fig.3-10}, the topological number is calculated as: $W=-1+1=0$.

Through the above calculations for $d=4,5,6,7$ cases, we show that the cases $d=4,5$ have the same topological number $W=+1$, which is different from the cases $d=6,7$ with $W=0$. Therefore, it implies that the dimension will affect on the topological number of the singly rotating Kerr black hole calculated by the R{\'e}nyi statistics, which is also mentioned in \cite{Bai:2022klw,Liu:2022aqt,Wu:2022whe,Wu:2023sue} for different kinds of black holes based on the GB statistics.

Now, we will also determine the topological number for general dimensional Kerr black holes via analyzing the asymptotic behaviors of $\tau_{\rm R}$ and explain the reason for the different topological numbers when $d>5$. We start with Eq.(\ref{kerrdtau}),
\begin{equation}
	\label{kerrdtau2}
	\tau_{\rm R}=\frac{16\pi r_{\rm h}^{5}((d-4)a^{2}+(d-2)r_{\rm h}^{2})}{(d-2)((d-3)r_{\rm h}^{2}+(d-5)a^{2})(4r_{\rm h}^{4}+\lambda  \omega _{d-2}r_{\rm h}^{d}(a^{2}+r_{\rm h}^{2}))}.
\end{equation}
Note that when $d=4$, the lower bound of $r_{\rm h}$ is $r_{\rm h} \to r_{\rm ex}=a$, thus the asymptotic
behavior of $\tau_{\rm R}$ is given by
\begin{equation}
	\begin{aligned}
		&\tau_{\rm R}\to \infty, \quad r_{\rm h}\to r_{\rm ex} ,\\
		&\tau_{\rm R}\to 0, \quad r_{\rm h}\to \infty ,
	\end{aligned}
\end{equation}
and the topological number $W$ is equal to $+1$. When $d=5$, the asymptotic behavior of $\tau_{\rm R}$ is same as the case $d=4$, but the lower bound of $r_{\rm h}$ is different, which is $r_{\rm h} \to r_{\rm min} =0$ in the small $r_{\rm h}$ limit. Therefore, the topological numbers are the same for both $d=4$ and $d=5$ cases.

For the $d>5$ case, we can find that the low bound of $r_{\rm h}$ in Eq.(\ref{kerrdtau2}) is zero (i.e., $r_{\rm h} \to r_{\rm min}=0$), hence the asymptotic behaviors of $\tau_{\rm R}$ in $r_{\rm h} \to 0$ and $r_{\rm h} \to \infty$ limits can be calculated as
\begin{equation}
	\begin{aligned}
		&\tau_{\rm R}\to 0, \quad r_{\rm h}\to 0 ,\\
		&\tau_{\rm R}\to 0, \quad r_{\rm h}\to \infty ,
	\end{aligned}
\end{equation}
which is the same as Type 4 in \cite{Liu:2022aqt} and the topological number is $W=0$.

In short, since the asymptotic behaviors of $\tau_{\rm R}$ for singly rotating Kerr black holes have an effect on the dimension, where there are the same asymptotic behavior for $d=4,5$, but there is a different asymptotic behavior for $d>5$, the topological number depends on the dimension. Based on the above calculation, we find that the topological numbers of cases $d=4,5$ and $d>5$ are equal to $+1$ and $0$, respectively, which is consistent with the results obtained from the GB statistics \cite{Wu:2022whe,Wu:2023sue}.

%%%%%%%%%%%%%%%%%%%%%%%%%%%%%%%%%%%%%%%%%%%%%%%%%%%%%%%%%%%%%%%%%%%%%%
\section{The topological number of five-dimensional Gauss-Bonnet black hole}
\label{5d GB black hole}
%%%%%%%%%%%%%%%%%%%%%%%%%%%%%%%%%%%%%%%%%%%%%%%%%%%%%%%%%%%%%%%%%%%%%%

\begin{figure}[htbp]
	\centering
	\subfloat[$\lambda <\lambda_{c} $]{\includegraphics[width=0.45\columnwidth,height=0.45\linewidth]{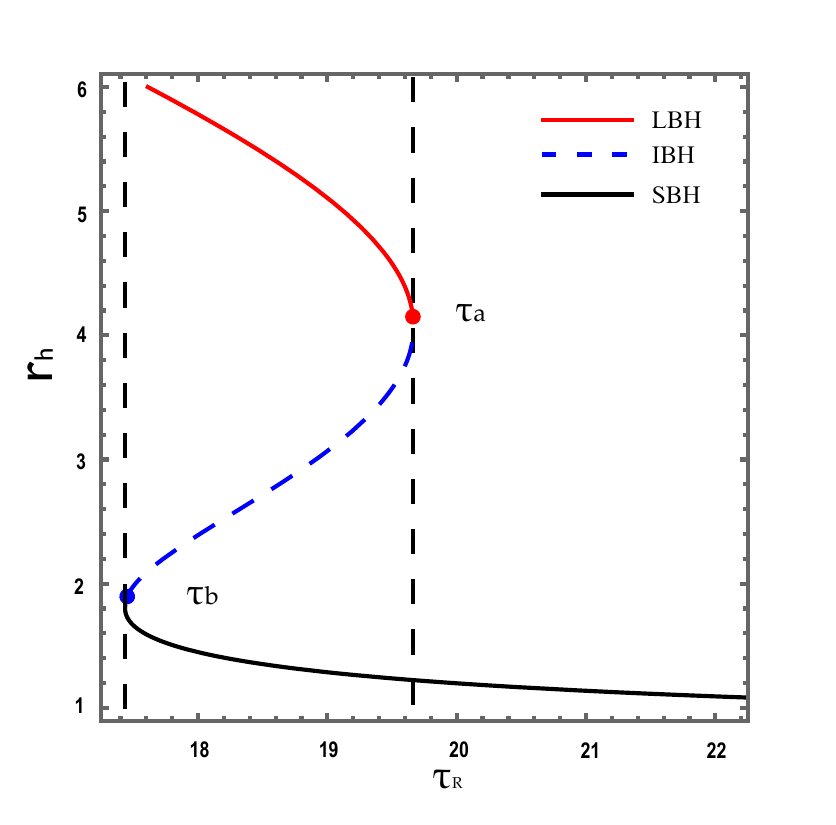}}
	\quad
	\subfloat[$\lambda > \lambda_{c} $]{\includegraphics[width=0.45\columnwidth,height=0.45\linewidth]{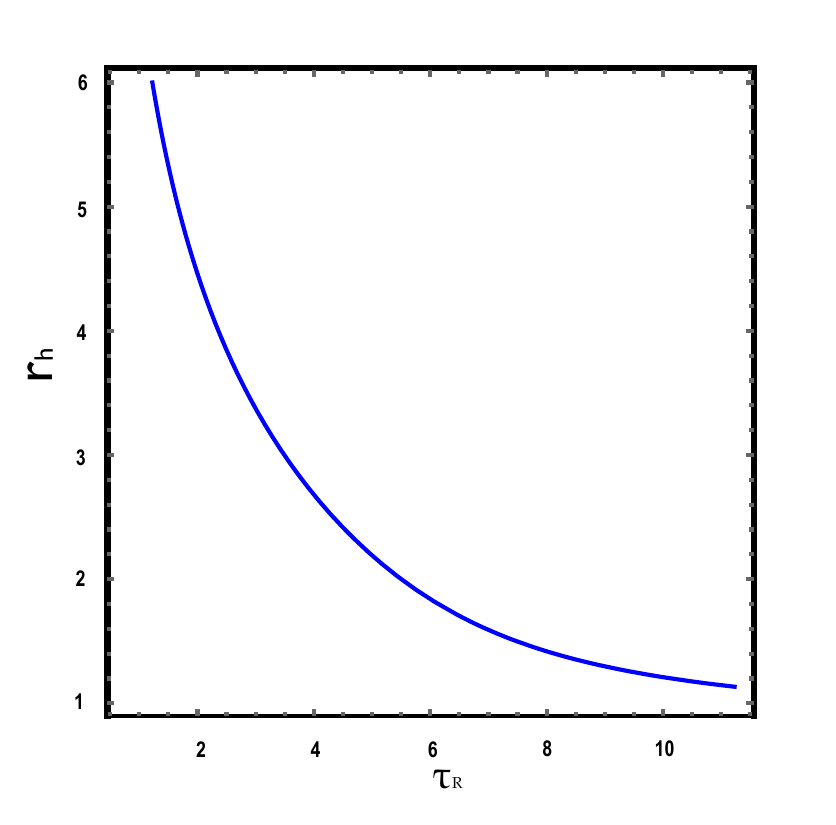}}
	
	\caption{The curves of equation (\ref{GBtauR}). The red solid, blue dashed, and black solid lines are for the large black hole (LBH), intermediate black hole (IBH), and small black hole (SBH), respectively. The annihilation and generation points are represented by red and blue dots, respectively. The left figure (a) is plotted with $Q=1$, $\alpha=1$ and $\lambda=0.001 <\lambda_{c}$. The right figure (b) is plotted with $Q=1$, $\alpha=1$, and $\lambda=0.1 >\lambda_{c}$.}
	\label{fig4-1ab}
\end{figure}

\begin{figure}[htbp]
	\centering
	\includegraphics[width=0.55\columnwidth,height=0.55\linewidth]{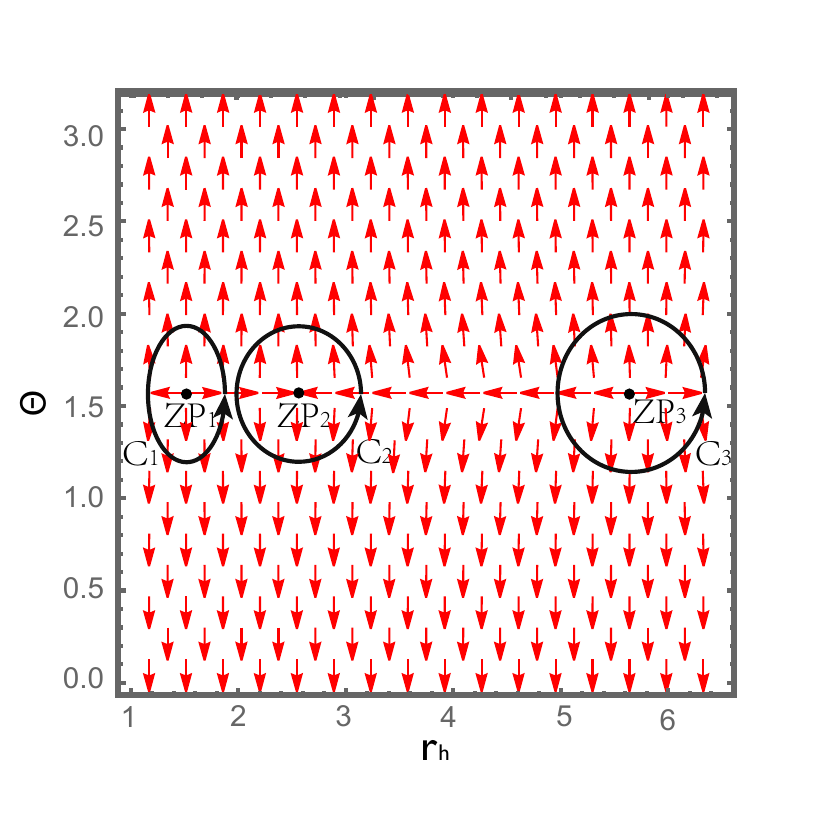}
	\caption{The red arrows represent the unit vector field $n$ on a portion of the $r_{\rm h}-\Theta$ plane with $\lambda = 0.001 $, $Q=1$, $\alpha=1$ and $\tau_{\rm R}=18$ for five-dimensional Gauss-Bonnet black hole. The black contours $C_{i}$ are closed loops surrounding
		the zero points.}
	\label{fig.4-3}
\end{figure}

In this section, we will study the topological numbers of five-dimensional Gauss-Bonnet black holes with and without charges, the $d$-dimensional spherically symmetric Gauss-Bonnet black hole is \cite{Cai:2001dz}
\begin{equation}
	ds^{2}=-f(r)dt^{2}+f(r)^{-1}dr^{2}+r^{2}d\Omega_{d-2}^{2} ,
\end{equation}
where $f(r)$ is
\begin{equation}
	\label{dGBmetric}
	f(r)=1+\frac{r^{2}}{2 \alpha}\left(1-\sqrt{1+\frac{64 \pi \alpha M}{(d-2) \omega _{d-2} r^{d-1}}-\frac{2 \alpha Q^{2}}{(d-2)(d-3) r^{2 d-4}}}\right),
\end{equation}
and $M$ is the ADM mass, $Q$ is the charge, and $\alpha$ is related to the positive Gauss-Bonnet coefficient $\alpha_{\rm GB}$, which satisfies $\alpha=(d-2)(d-3)\alpha_{\rm GB}$. For $d=5$, the metric in Eq.(\ref{dGBmetric}) reduces to
\begin{equation}
	f(r)=1+\frac{r^{2}}{2 \alpha }\left(1-\sqrt{1+\frac{64 \pi\alpha  M}{3\omega _{3}  r^{4}}-\frac{  \alpha Q^{2}}{3 r^{6}}}\right).
\end{equation}
We list the thermodynamic quantities as follows:
\begin{equation}
	M  = \frac{\omega _{3} Q^{2}}{64\pi r_{\rm h}^{2}}+\frac{3 \omega _{3} r_{\rm h}^{2}}{16\pi }\left(1+\frac{\alpha}{r_{\rm h}^{2}} \right), \quad S_{\rm BH}=\frac{\omega _{3} r_{\rm h}^{3}}{4}\left(1+\frac{6\alpha}{r_{\rm h}^{2}} \right).
\end{equation}
Thus, we obtain the R{\'e}nyi entropy as
\begin{equation}
	S_{\rm R}=\frac{1}{\lambda}\ln\left(1+\lambda \frac{\omega _{3} r_{\rm h}^{3}}{4}\left(1+\frac{6\alpha}{r_{\rm h}^{2}} \right)\right) ,
\end{equation}
and the generalized free energy of the five-dimensional charged Gauss-Bonnet black hole can be written as
\begin{equation}
	\mathcal{F}_{\rm R}=\ \frac{\omega _{3} Q^{2}}{64\pi r_{\rm h}^{2}}+\frac{3 \omega _{3} r_{\rm h}^{2}}{16\pi }\left(1+\frac{\alpha}{r_{\rm h}^{2}} \right)-\frac{1}{\lambda \tau_{\rm R}}\ln\left(1+\lambda \frac{\omega _{3} r_{\rm h}^{3}}{4}\left(1+\frac{6\alpha}{r_{\rm h}^{2}} \right)\right).
\end{equation}
The components of the vector $\phi$ can be calculated as
\begin{equation}
	\begin{aligned}
		\phi ^{r_{\rm h}}&=\frac{\omega_{3} (12r_{\rm h}^{4}-Q^{2})}{32\pi r_{\rm h}^{3}}-\frac{3 \omega_{3} (2\alpha+r_{\rm h}^{2})}{(4+\lambda \omega_{3} r_{\rm h}(6\alpha+r_{\rm h}^{2}))\tau_{\rm R}},     \\
		\phi ^{\Theta }&=-\cot \Theta \csc \Theta.
	\end{aligned}
\end{equation}
By solving the equation $\phi^{r_{\rm h}}=0$, we obtain
\begin{equation}
	\label{GBtauR}
	\tau_{\rm R}=\frac{96\pi (2 \alpha r_{\rm h}^{3}+r_{\rm h}^{5})}{(12r_{\rm h}^{4}-Q^{2})(4+\lambda \omega_{3} r_{\rm h}(6\alpha+r_{\rm h}^{2}))}.
\end{equation}
It is worth noting that through the analysis of Eq.(\ref{GBtauR}), it can be found that there is a critical value $\lambda_{c}$. when choosing $Q=1$ and $\alpha=1$, we can find $\lambda_{c}\simeq 0.00229$ numerically. As shown in Fig.\ref{fig4-1ab}, we choose $Q=1$, $\alpha=1$ and plot the curve of Eq.(\ref{GBtauR}) for fixing $\lambda=0.001<\lambda_{c}$ and $\lambda=0.1>\lambda_{c}$  in $r_{\rm h}-\tau_{\rm R}$ plane. In Fig.\ref{fig.4-3}, we show the unit vector $n$ with $\tau_{\rm R}=18$, there are three zero points which are located at ZP$_i$ $(r_{\rm h},\Theta)=(1.20,\pi/2), \ (2.47,\pi/2), \ (5.85,\pi/2)$. Thus, the topological number is calculated as: $W=+1-1+1=+1$.

In the following part, we focus on the analysis of asymptotic behaviors of $\tau_{\rm R}$ and calculate the topological numbers when the charge is absent. Besides, as a comparison, we also use the analysis of asymptotic behavior of $\tau_{\rm R} $ to give the topological numbers of five-dimensional Gauss-Bonnet black hole calculated by GB statistics. When $Q=0$, Eq.(\ref{GBtauR}) reduces to
\begin{equation}
	\label{unchargedGBtauR}
	\tau_{\rm R}=\frac{8\pi (2\alpha+r_{\rm h}^{2})}{r_{\rm h}(4+\lambda \omega_{3} r_{\rm h}(6\alpha+r_{\rm h}^{2}))},
\end{equation}
then the lower bound of $r_{\rm h}$ in Eq.(\ref{unchargedGBtauR}) is zero, hence the asymptotic behavior of $\tau_{\rm R}$ is given by
\begin{equation}
	\begin{aligned}
		&\tau_{\rm R}\to \infty, \quad r_{\rm h}\to 0 ,\\
		&\tau_{\rm R}\to 0, \quad r_{\rm h}\to \infty ,
	\end{aligned}
\end{equation}
thus the topological number is $W=+1$, which is same as the case of $Q \ne 0$.

Now taking the limit of $\lambda \to 0$, the entropy function $S_{\rm R}$ reduces to $S_{\rm BH}$ and $\tau_{\rm R}$ becomes
\begin{equation}\label{tauGBGB}
	\tau=	\frac{24\pi (2 \alpha r_{\rm h}^{3}+r_{\rm h}^{5})}{12r_{\rm h}^{4}-Q^{2}},
\end{equation}
which exists a lower bound of $r_{\rm ex}$. It is easy to find that $\tau$ possesses the asymptotic behaviors:
\begin{equation}
	\begin{aligned}
		&\tau\to \infty, \quad r_{\rm h}\to r_{\rm ex} ,\\
		&\tau\to \infty, \quad r_{\rm h}\to \infty ,
	\end{aligned}
\end{equation}
which implies the topological number is $W=0$. In addition, when setting $Q=0$ in Eq.(\ref{tauGBGB}), the asymptotic behaviors of $\tau$ are the same as the $Q\neq 0$ case. Thus, the topological number is $W=0$ both for the five-dimensional charged and uncharged Gauss-Bonnet black holes calculated by GB statistics, which is different from the topological number $W=+1$ obtained by the R{\'e}nyi statistics.
%%%%%%%%%%%%%%%%%%%%%%%%%%%%%%%%%%%%%%%%%%%%%%%%%%%%%%%%%%%%%%%%%%%%%%
\section{Conclusions and discussions}\label{discussion}
%%%%%%%%%%%%%%%%%%%%%%%%%%%%%%%%%%%%%%%%%%%%%%%%%%%%%%%%%%%%%%%%%%%%%%

In this paper, we combined the topological method with the R{\'e}nyi statistics to investigate black hole thermodynamics, and calculated the topological number of various black holes, the results are listed in Tab.\ref{tab.1}, from which one can see that the topological numbers calculated from the R{\'e}nyi statistics are different from previous results calculated via the GB statistics, it is natural because $\tau_{\rm R}$ is different from $\tau$. However, although the topological numbers are different between the calculations of R{\'e}nyi and GB statistics, we find that the topological classifications of various types of vacuum black holes share similar properties as their counterparts in AdS spacetime: four-dimensional RN, four-dimensional and five-dimensional singly rotating Kerr, five-dimensional charged and uncharged Gauss-Bonnet black holes belong to the same kind of topological classes, and four-dimensional Schwarzschild black hole and $d>5$ singly rotating Kerr black holes belong to the another different kind of topological classes.
\begin{table}[htbp]
	\renewcommand{\arraystretch}{1.2}
	\centering
	\caption{The topological number $W$ calculated from the R{\'e}nyi and the GB statistics for different black holes.}
	\setlength{\tabcolsep}{4mm}{
		\begin{tabular}{|c|c|c|}
			\hline
			Black holes (BHs)                                                                                         & \begin{tabular}[c]{@{}c@{}}Topological number \\ (R{\'e}nyi statistics)\end{tabular} & \begin{tabular}[c]{@{}c@{}}Topological number \\ (GB statistics)\end{tabular} \\ \hline
			\begin{tabular}[c]{@{}c@{}}$d=4$ RN BH \\ $d=4$ Kerr BH\\ $d=5$ singly rotating Kerr BH\\$d=5$ charged Gauss-Bonnet BH \\ $d=5$ uncharged Gauss-Bonnet BH \end{tabular} & $W=+1$                                                                                       & $W=0$                                                                           \\ \hline
			\begin{tabular}[c]{@{}c@{}}$d=4$ Schwarzschild BH\\ $d>5$ singly rotating Kerr BH\end{tabular}      & $W=0$                                                                                        & $W=-1$                                                                          \\ \hline
	\end{tabular}}
	\label{tab.1}
\end{table}

Furthermore, an interesting point we mentioned at the end of Sec.\ref{Schwarzschild topology} is that the topological numbers in asymptotically flat spacetime calculated via the R{\'e}nyi statistics seem to be related to those of black holes in asymptotically AdS spacetime calculated via the GB statistics. As shown in Tab.\ref{tab.2}, by comparing the topological numbers we calculated with previous results obtained from the GB statistics, we found that their topological numbers can correspond with each other. Note that there were many studies focusd on black holes thermodynamics via the R{\'e}nyi statistics, and showed that the nonextensive parameter $\lambda$ plays the role of pressure just like the cosmological constant $\Lambda$ in AdS background. Thus, it is natural that we can obtain the same topological number as in AdS spacetime via the R{\'e}nyi statistics. In fact, from another point of view, we utilized the topological method to provide an evidence for the connection between the R{\'e}nyi parameter $\lambda$ and the cosmological constant $\Lambda$. Especially, previous studies on black hole thermodynamics analyzed via the R{\'e}nyi statistics mainly focused on four-dimensional cases (e.g., Schwarzschild, RN, Kerr, etc.), in the present paper, we used the topological method to calculate not only the four-dimensional case, but also the higher-dimensional black holes and five-dimensional Gauss-Bonnet black hole. We showed that the topological numbers calculated from R{\'e}nyi and GB statistics in asymptotically flat and AdS spacetimes are also consistent, respectively, which reveals that the thermodynamic behaviors of higher dimensional black holes and Gauss-Bonnet gravity in asymptotically flat spacetime via the R{\'e}nyi statistics is connected with their counterparts in asymptotically AdS spacetime via the GB statistics. Our results may shed light on future studies on the topological classes of many other types of black holes via the R{\'e}nyi statistics and also the entanglement entropy of black holes. Another interesting problem to study is whether the topological numbers of black holes can be understood from the black 
hole shadow obtained by the Event Horizon Telescope, we would like to study this problem in a future work.

\begin{table}[htbp]
	\renewcommand{\arraystretch}{1.2}
	\centering
	\caption{The topological number $W$ calculated via  R{\'e}nyi and GB statistics for different black holes.}
	\setlength{\tabcolsep}{5mm}{
		\begin{tabular}{|c|c|c|}
			\hline
			Black holes                                                                                               & \begin{tabular}[c]{@{}c@{}}Topological \\ number $W$\end{tabular} & Statistics                                    \\ \hline
			\begin{tabular}[c]{@{}c@{}}$d=4$ Schwarzschild BH\\ Schwarzschild-AdS$_{4}$ BH \cite{Wu:2023sue}\end{tabular}               & 0                                                                 & \begin{tabular}[c]{@{}c@{}}R{\'e}nyi statistics\\ GB statistics\end{tabular} \\ \hline
			\begin{tabular}[c]{@{}c@{}}$d$-dimensional RN BH\\ RN-AdS$_{d}$ BH \cite{Wei:2022dzw}\end{tabular}                           & +1                                                                & \begin{tabular}[c]{@{}c@{}}R{\'e}nyi statistics\\ GB statistics\end{tabular} \\ \hline
			\begin{tabular}[c]{@{}c@{}}$d=4$ Kerr BH \\ Kerr-AdS$_{4}$ BH \cite{Wu:2023sue}\end{tabular}                                 & +1                                                                & \begin{tabular}[c]{@{}c@{}}R{\'e}nyi statistics\\ GB statistics\end{tabular} \\ \hline
			\begin{tabular}[c]{@{}c@{}}$d=5$ singly rotating Kerr BH\\  singly rotating Kerr-AdS$_{5}$ BH \cite{Wu:2023sue}\end{tabular} & +1                                                                & \begin{tabular}[c]{@{}c@{}}R{\'e}nyi statistics\\ GB statistics\end{tabular} \\ \hline
			\begin{tabular}[c]{@{}c@{}}$d>5$ singly rotating Kerr BH\\ $d>5$ singly rotating Kerr-AdS BH \cite{Wu:2023sue}\end{tabular} & 0                                                                 & \begin{tabular}[c]{@{}c@{}}R{\'e}nyi statistics\\ GB statistics\end{tabular} \\ \hline
			\begin{tabular}[c]{@{}c@{}}$d=5$ charged Gauss-Bonnet BH\\  charged Gauss-Bonnet-AdS$_{5}$ BH \cite{Liu:2022aqt}\end{tabular} & +1                                                                 & \begin{tabular}[c]{@{}c@{}}R{\'e}nyi statistics\\ GB statistics\end{tabular} \\ \hline
			\begin{tabular}[c]{@{}c@{}}$d=5$ uncharged Gauss-Bonnet BH\\ uncharged Gauss-Bonnet-AdS$_{5}$ BH \cite{Liu:2022aqt}\end{tabular} & +1                                                                 & \begin{tabular}[c]{@{}c@{}}R{\'e}nyi statistics\\ GB statistics\end{tabular} \\ \hline
	\end{tabular}}
	\label{tab.2}
\end{table}

%%%%%%%%%%%%%%%%%%%%%%%%%%%%%%%%%%%%%%%%%%%%%%%%%%%%%%%%%%%%%%%%%%%%%%
\section*{Acknowledgement}
%%%%%%%%%%%%%%%%%%%%%%%%%%%%%%%%%%%%%%%%%%%%%%%%%%%%%%%%%%%%%%%%%%%%%%
We would like to thank S.-W. Wei for useful discussions. This work was supported by the National Natural Science Foundation of China (No.~11675272).

%%%%%%%%%%%%%%%%%%%%%%%%%%%%%%%%%%%%%%%%%%%%%%%%%%%%%%%%%%%%%%%%%%%%%%
%\begin{appendix}
%%%%%%%%%%%%%%%%%%%%%%%%%%%%%%%%%%%%%%%%%%%%%%%%%%%%%%%%%%%%%%%%%%%%%%
%\section*{APPENDIX}

%\subsection*{APPENDIX A}

%\end{appendix}

%%%%%%%%%%%%%%%%%%%%%%%%%%%%%%%%%%%%%%%%%%%%%%%%%%%%%%%%%%%%%%%%%%%%%%
%\begin{references}{999}
	
%\bibliography{ref}
\end{document}